\documentclass[11pt]{article}
\usepackage{graphicx}
\usepackage[table,xcdraw]{xcolor}
\usepackage{lineno}
\usepackage{optidef}
\usepackage{lscape}
\usepackage{booktabs}
\usepackage{multirow}
\usepackage{hyperref}
\hypersetup{colorlinks,allcolors=black}
\usepackage[flushleft]{threeparttable}
\usepackage{subcaption}
\usepackage{natbib}
\usepackage[superscript]{cite}
\usepackage{authblk}

\begin{document}

\title{Evaluation of Phase I Clinical Trial Designs for Combinational Agents along with Guidance based on Simulation Studies}

\author[1,2]{Shu Wang}
\author[3]{Elias Sayour}
\author[1,2]{Ji-Hyun Lee}
\affil[1]{\small{Division of Quantatitive Sciences, UF Health}}
\affil[2]{\small{Department of Biostatistics, University of Florida}}
\affil[3]{\small{Department of Neurosurgery, University of Florida}}

\renewcommand\Authands{ and }

\maketitle

\begin{abstract}
Combinational therapy that combines two or more therapeutic agents is very common in cancer treatment. Currently, many clinical trials aim to assess feasibility, safety and activity of combinational therapeutics to achieve synergistic response. Dose-finding for 
combinational agents is considerably more complex than single agent, because only partial order of dose combinations' toxicity is known. Prototypical phase I designs 
may not adequately capture this complexity thus limiting identification of the maximum 
tolerated dose (MTD) of combinational agents. In response, novel phase I 
clinical trial designs for combinational agents have been extensively proposed. However, with so many available designs, studies that compare their performances and explore the impact of design parameters, along with providing recommendations are limited. 
We are evaluating available phase I designs that identify a single MTD for combinational agents using simulation studies under various conditions. We are also exploring the influences of different design parameters and summarizing the risks/benefits of each design to provide general guidance in design selection.

\textit{KEY WORDS:} clinical trial, combinational agents, dose-finding, phase I
\end{abstract}

\section{Introduction}\label{sec1}

Clinical trials investigating combinational therapies that combine two or more therapeutic agents have garnered renewed attention with the development of cancer therapy.  
Novel combinational agents require identification of a maximum tolerated dose (MTD). The MTD is the highest dose level that leads to a pre-specified target toxicity probability.
However, dose-finding for combinational agents could be challenging as we do not know the complete order of dose combinations' toxicity. 
To fill the gap, several phase I study designs for combinational agents have been proposed.

Overall, there are 3 categories of designs: algorithm- or rule-based, model-based, and model-assisted. 
Algorithm-based designs do not involve any parametric relationship between dose combinations and their toxicity probabilities. Ivanova and Wang \citep{ivanova2004non} proposed an up-and-down design and used isotonic regression to estimate the MTD. Later Ivanova and Kim\citep{ivanova2009dose} updated the previous up-and-down design using T-statistics. Lee and Fan \citep{lee2012two} proposed a two-dimensional search algorithm to identify MTD. 
Algorithm-based designs usually lack statistical theory foundation, and their escalation/de-escalation rules are ad-hoc. Therefore, their performances are not guaranteed.

Model-based designs assume a parametric dose-toxicity relationship. 
To account for the design parameters' estimation uncertainty in the beginning of the trial, a start-up phase is usually used before transitioning to model-based part. 
The main differences among model-based methods are the choice of dose-toxicity relationship and the scheme of start-up phase. Thall et al. \citep{thall2003dose} proposed to identify MTD contour with a six-parameter logistic regression. 
Wang and Ivanova \citep{wang2005two} proposed a three parameter model to link doses and toxicity probabilities. 
Yin and Yuan proposed a latent contingency table method \citep{yin2009latent} and another one that used Copula to model toxicity probabilities through marginal toxicity profile of individual agents \citep{yin2009bayesian}. 
Riviere et al. \citep{riviere2014bayesian} developed a method based on Bayesian logistic regression. 
Braun and Jia \citep{braun2013generalized}  used a proportional odds logistic regression fitting model within each ``row" of the dose combination matrix, and later join them together. 
Braun and Wang \citep{braun2010hierarchical} proposed a hierarchical model through linking the effective doses with hyperparameter of dose toxicity probabilities. 
Tighiouart et al. \citep{tighiouart2017bayesian} extended the escalation with overdose control (EWOC) method \citep{babb1998cancer} to two-dimensional setting to identify the MTD curve. 
Some other designs were motivated by the main difficulty of two-dimensional dose-finding where only the partial order of dose combinations' toxicity is known. In this case, we only know that doses $(i+1,j)$ and $(i,j+1)$ are more toxic than dose $(i,j)$, but we are not aware of the toxicity order of dose $(i-1,j+1)$ and $(i,j)$, where $(i,j)$ denotes the dose combination of the $i^{th}$ dose of one agent and the $j^{th}$ dose of the second agent.
Therefore, several possible toxicity orderings which satisfy the partial order exist.
Conaway et al. \citep{conaway2004designs} proposed to identify all possible orderings of dose combination toxicities so that the two dimensional dose-finding can be solved by continuous reassessment method (CRM) \citep{o1990continual}. 
Building upon this methodology, Wages et al. \citep{wages2011continual,wages2011dose} proposed to use a subset of possible orderings, which is more feasible especially when the total number of possible orderings is large. 
To avoid pre-specifying orderings, Lin and Yin \citep{lin2016bootstrap} proposed to dynamically update the ordering. 
However, model-based designs have several limitations in practice: (1) they are relatively complicated and require constant model updating by statisticians, which posit barriers to clinicians to understand and implement. 
(2) Most designs need parameter calibration. 
(3) Some designs require prior knowledge about agents (e.g., guesses of dose combinations' toxicities) which is not easy for clinicians to provide.   

Model-assisted designs still utilize statistical models in decision making, but focus on easier implementation through pre-tabulating escalation and de-escalation rules before trial conduct \citep{yan2017keyboard}. Therefore, they have the advantages of algorithm-based and model-based designs. 
BOIN design \citep{liu2015bayesian,yuan2016bayesian} and Keyboard design \citep{yan2017keyboard} are representative model-assisted designs.
Later, Lin and Yin \citep{lin2017bayesian} extended the BOIN design, Pan et al. \citep{pan2020keyboard} extended the Keyboard design to handle two-dimensional dose-findings.

In addition, there are designs that incorporate special features while conducting two dimensional dose-finding. Liu and Ning \citep{liu2013bayesian} proposed a design that is able to handle trials with delayed toxicities. Diniz et al. \citep{diniz2018bayesian} built a Bayesian design for combinational doses upon escalation with the EWOC method \citep{babb1998cancer} accounting for patient heterogeneity through taking baseline covariates into consideration. 

With so many novel methods available yet very limited implementation in oncology trials, we have relatively little knowledge about which methods are superior under which scenarios. 
Riviere et al. \citep{riviere2015competing} compared six phase I designs for combinational agents that are either algorithm-based or model-based. Their paper claimed that all designs were optimized to improve the percentage of correct MTD selection before comparison. 
However, such claim itself is questionable as it is impossible to achieve optimization under diverse scenarios using a universal set of parameters. Based on the sensitivity analyses the authors have conducted, different scenarios actually required different parameter sets to achieve optimization. 
Therefore, simply comparing results using single set of design parameters makes the comparison less meaningful. Another limitation of this study is that it did not discuss in detail about the influences of different design settings, although sensitivity analyses were presented.
Hirakawa et al. \citep{hirakawa2015comparative} compared performances of five model-based designs for combinational agents. But this paper did not explore effects of design parameter beyond cohort size.
Harrington et al. \citep{harrington2013adaptive} reviewed some algorithm-based and model-based combinational agents designs, and discussed their advantages and limitations without simulation studies.
None of the above papers included the recently developed model-assisted designs as they were published in earlier days. Moreover, the feasibility of parameter tuning and the influences of design parameters for the model-based designs were investigated in a limited fashion.
Later Pan et al. \citep{pan2020keyboard} compared two-dimensional BOIN, two-dimensional Keyboard, and Continual reassessment method for partial ordering (POCRM) \citep{wages2011continual,wages2011dose}. However, POCRM was the only model-based design in the comparison. Moreover, similar to those review studies mentioned above, this one did not explore effects of design parameters for POCRM.
To provide a more recent view of phase I clinical trial designs for combinational agents, we conducted a simulation study to evaluate the performances of various designs under comprehensive scenarios. 

Our study is different from previous review papers in several aspects: (1) we included two recently-developed model-assisted designs in the study, (2) for design parameters with no clear recommendations, we investigated multiple sets to investigate their influences instead of using single subjectively selected set, (3) we used different sample sizes in simulations to check that whether our findings are valid with different trial sizes, and (4) in addition to the summary of each design's characteristics, we discussed putative reasons that led to their performances. 

Specifically, in this paper we focus on the designs that (1) utilize toxicity information only, (2) identify single MTD instead of MTD contour or curve, (3) assume monotonic dose-toxicity relationship within each individual agent, and (4) having programming codes/softwares available. 
As a result, we have selected 9 designs: Dose finding in discrete dose space \citep{wang2005two}, Bayesian dose finding by copula regression \citep{yin2009bayesian}, Continual reassessment method for partial ordering \citep{wages2011continual,wages2011dose}, Hierarchical Bayesian Design \citep{braun2010hierarchical}, Logistic model-Based Bayesian dose finding design \citep{riviere2014bayesian}, Generalized Continual Reassessment Method \citep{braun2013generalized}, Bootstrap Aggregating Continual Reassessment Method \citep{lin2016bootstrap}, combinational Bayesian optimal interval design \citep{lin2017bayesian}, and combinational Keyboard design \citep{pan2020keyboard}. 
We did not include few currently available algorithm-based designs as it is consensus that algorithm-based designs usually have inferior performances compared with model-based ones \citep{korn1993using,riviere2015competing,love2017embracing}. 

The rest of the paper is organized as follows: we first reviewed nine designs that will be included in our evaluation; then presented our simulation studies and results; in the last, we discussed our findings.

\section{Review of Designs}
\label{sec:p2}

\subsection{Notations}
Here we define some common notations used in these methods. As most of the designs we selected apply to dual-agent dose finding only (Copula has been extended to handle more than two agents), we assume two agents $A$ and $B$, with $J$ and $K$ doses respectively. Define $\pi_{jk}$ to be the true toxicity probability of the dose combination $(j,k)$, $j=1,2,\dots, J$, $k=1,2,\dots, K$; define $p_j$ to be the true toxicity probability of agent $A$ when used as a monotherapy, $j=1,2,\dots, J$, and $q_k$ to be the true toxicity probability of agent $B$ when used as a monotherapy, $k=1,2,\dots, K$. 
Define $\phi$ to be the pre-specified target toxicity probability. 
Define $N$ to be maximum sample size in the trial. Define $n_{jk}$ to be number of subjects that received dose $(j,k)$ and $y_{jk}$ to be number of dose-limiting toxicities (DLTs) observed among those $n_{jk}$ patients.

\subsection[I2D]{Model-based: Dose finding in discrete dose space (I2D)}
This method is a Bayesian design that extends CRM to accommodate dose-finding in dual agents. 

\textbf{Dose-toxicity model}:
\begin{equation} \label{eq1}
\pi_{jk}(\mathbf{\theta})=1-(a_j)^\alpha(1-b_k)^{\beta+\gamma\log(1-a_j)},
\end{equation}
where $\mathbf{\theta}=(\alpha,\beta,\gamma)$ is a vector of unknown design parameters and restricts $\alpha>0, \beta>0, \gamma<0$ to satisfy the assumption of toxicity monotonicity; $0\leq a_1<\dots<a_J$ and $0\leq b_1<\dots<b_K$ are constants instead of actual doses of agents. 
 If no interaction between two agents exists in Equation~\ref{eq1}, the model becomes
\begin{equation} \label{eq2}
\pi_{jk}(\mathbf{\theta})=1-(a_j)^\alpha(1-b_k)^\beta.
\end{equation}

\textbf{Start-up phase}:
the trial is initiated with a dose combination $(1,1)$. 
Next, escalate agent $A$ while agent $B$ is maintained at the lowest dose if no DLT is observed. 
If still no DLT is observed when agent $A$ reaches its maximum dose, agent $B$ is escalated to its second lowest dose combining with agent $A$'s $(J-2)^{th}$ dose. 
Then if no DLT is observed, we continue to a combination where agent $B$ is at its third lowest dose and agent $A$'s $(J-4)^{th}$ dose, namely the one used in previous combination minus 2.
When agent $B$ reaches its maximum dose, if agent $A$ is at its $m^{th}$ dose, we evaluate all combinations from $(m,K), (m+1,K),\dots,(J,K)$. The start-up phase ends if at least one DLT is observed at any time.

\textbf{Post start-up escalation / de-escalation dose set}:
if the current combination is $(j,k)$, I2D only considers doses $(j-1,k),(j+1,k),(j,k),(j,k+1),(j,k-1),(j+1,k-1),(j-1,k+1)$ to the next subject prohibiting diagonal moves.

\textbf{Post start-up trial conduct}:
after the start-up phase ends, the working model Equation~\ref{eq1} or Equation~\ref{eq2} will be used to obtain toxicity estimates of all dose combinations. 
Due to safety concerns, the working model starts at the combination dose where agent $B$ is at its lowest dose and agent $A$ is at the dose that makes the combination's estimated toxicity probability closest to $\phi$. 

\textbf{MTD determination}:
the dose combination whose posterior probability of toxicity is closest to $\phi$ will be selected as the MTD.

\subsection[Copula]{Model-based: Bayesian dose finding by copula regression (Copula)}
This method utilizes copula to model the dose-toxicity relationship because copula allows to link the joint distribution and marginal distributions via a dependence parameter. 

\textbf{Dose-toxicity model}:
\begin{equation} \label{eq3}
\pi_{jk}=1-\{(1-p_j^\alpha)^{-\gamma}+(1-q_k^\beta)^{-\gamma}-1\}^{-1/\gamma},
\end{equation}
where $\alpha$ and $\beta$ are power parameters as in CRM to accommodate the uncertainty, and $\gamma>0$ represents the interaction between two agents. Intermediate informative prior distributions with prior mean 1 and a relatively small variance will be assigned to $\alpha$ and $\beta$ (e.g. Gamma (2,2)). A Gamma distribution with a large variance is usually chosen as the non-informative prior for $\gamma$.
If only one agent is involved, this approach reduces to regular CRM.

\textbf{Start-up phase}:
the start-up phase begins with the lowest dose combination $(1,1)$. It proceeds vertically $(1,2), (1,3),\dots,(1,K)$ until the first toxicity is observed, then it proceeds horizontally $(2,1), (3,1),\dots,(J,1)$ until the first toxicity is observed. 
Once one toxicity is observed in both directions, the formal design starts.

\textbf{Post start-up escalation / de-escalation dose set}:
if the current combination is $(j,k)$, the dose escalation set is defined as ${\cal A}_E=\{(j+1,k),(j,k+1),(j+1,k-1),(j-1,k+1)\}$, dose de-escalation set is define to be ${\cal A}_D=\{(j-1,k),(j,k-1),(j+1,k-1),(j-1,k+1)\}$. 
As we only know the partial order of dose toxicity in combinational agents, we do not know whether dose combinations $(j+1,k-1)$ and $(j-1,k+1)$ are more/less toxic than dose combination $(j,k)$. Therefore, the authors included $(j+1,k-1)$ and $(j-1,k+1)$ in both escalation and de-escalation sets. 

\textbf{Post start-up trial conduct}:
this design involves two parameters $c_e$ and $c_d$ that represent the fixed probability cut-offs for dose escalation and de-escalation, respectively, and $c_e+c_d>1$. 
Detailed algorithm is laid out as below. 
\renewcommand{\labelenumi}{\alph{enumi}}
\begin{enumerate}
    \item if at current dose combination $(j,k)$, $P(\pi_{jk}<\phi)>c_e$, we will escalate to the dose that belongs to ${\cal A}_E$ and with the toxicity probability closest to $\phi$ and higher than that of $(j,k)$. If current dose is $(J,K)$, then stay at the same combination.
    \item If at current dose combination $(j,k)$, $P(\pi_{jk}>\phi)>c_d$, we will de-escalate to the dose that belongs to ${\cal A}_D$ and with the toxicity probability closest to $\phi$ and lower than that of $(j,k)$. If current dose is $(1,1)$, the trial is terminated.
    \item Otherwise, stays at the same dose combination. 
\end{enumerate}

\textbf{MTD determination}:
after $N$ subjects are exhausted, the MTD is determined as the dose combination with the estimated probability of toxicity closest to $\phi$.

\subsection[POCRM]{Model-based: Continual reassessment method for partial ordering (POCRM)}
To solve the problem of only knowing partial order of dose toxicity, POCRM proposes to pre-specify a subset of possible orderings, then utilize the CRM on each of them. 
This way, two-dimensional dose-finding is reduced to a one-dimensional problem.  

\textbf{Dose-toxicity model}:
define $T$ as the total number of dose combinations, $T=J\times K$; $d_n$ as the dose assigned to subject $n$; $R(d_n)$ as the true toxicity probability of $d_n$; $y_n$ as a binary indicator of whether subject $n$ has toxicity or not, $n=1,2,\dots,N$; and $\Omega_n$ as data collected after having $n$ subjects where $\Omega_n=\{d_1,y_1,\dots,d_n,y_n\}$. 
Assume we have $M$ possible partial ordering in total. For a specific ordering $m$, $m=1,2,\dots,M$, $R(d_n)$ is modeled as below, similar to the CRM
\begin{equation} \label{eq4}
R(d_n)=E(Y_n|d_n)\doteq \psi_m(d_n,a)
\end{equation}
where $\psi_m$ is some working dose-toxicity model, $a$ is the model parameter. After having $n$ patients, the likelihood under partial order $m$ is
\begin{equation} \label{eq5}
L_m(a|\Omega_n)=\prod_{i=1}^{n} \psi_m^{y_i}(d_i,a)\{1-\psi_m(d_i,a)\}^{(1-y_i)}.
\end{equation}
Then, the estimate of parameter $a$ under ordering $m$, $\hat{a_m}$, could be obtained through maximizing Equation~\ref{eq5}. Then we could obtain the posterior probability of partial order $m$:
\begin{equation} \label{eq6}
p(m|\Omega_n)=\frac{p(m)L_m(\hat{a_m}|\Omega_n)}{\sum_{l=1}^{M} p(l) L_l(\hat{a_l}|\Omega_n)}.
\end{equation}

\textbf{Start-up phase}:
when POCRM was first proposed, it was a single stage method \citep{wages2011continual}. Later it was extended to include a start-up phase \citep{wages2011dose}. 
The start-up phase partitions the dose combination matrix to different ``zones" and starts the first cohort with zone 1, which is the lowest dose combination. 
If no DLT is observed, then it assigns next cohort to doses in zone 2. 
If there are multiple combinations in zone 2, it randomly selects one of them. If no DLT is observed, it continues to assign the next cohort to other combinations in the same zone. Moving to the next zone is only allowed when all the dose combinations have been explored in lower zones. 
The start-up phase ends when one DLT is observed. POCRM also allows users to specify their own scheme in the start-up phase as they see appropriate.

\textbf{Post start-up escalation / de-escalation dose set}:
since POCRM pre-specifies a subset of possible orderings, the escalation and de-escalation dose are known within each ordering.

\textbf{Post start-up trial conduct}:
after the start-up phase ends, the authors used weighted randomization to select the partial ordering with $p(m|\Omega_n)$ from Equation~\ref{eq6} being the weight.
After selecting the working partial ordering $m$, $\pi_t$ is estimated for all $t\in \{1,2,\dots,T\}$ through Equation~\ref{eq4} and assign the dose combination that minimizes $|\hat{\pi_t}-\phi|$ to the next subject.
But for final MTD determination, the ordering with the maximum posterior probability will be chosen among all candidate orderings.

\textbf{MTD determination}:
after $N$ subjects are exhausted, the MTD is determined as the dose combination with the estimated probability of toxicity closest to $\phi$, given the ordering with maximum posterior probability.

\subsection[Hierarchy]{Model-based: Hierarchical Bayesian Design (Hierarchy)}\label{hierarchy}

\textbf{Dose-toxicity model}:
This method employs a hierarchical model:
\begin{equation} \label{eq7}
\pi_{jk}\sim Beta(\alpha_{jk},\beta_{jk}),
\end{equation}
\begin{equation} \label{eq8}
log(\alpha_{jk})=\theta_0+\theta_1a_j+\theta_2b_k,
\end{equation}
\begin{equation} \label{eq82}
log(\beta_{jk})=\phi_0+\phi_1a_j+\phi_2b_k,
\end{equation}
where $\mathbf{\theta}=\{\theta_0,\theta_1,\theta_2\}$ follows a multivariate normal distribution with mean $\mathbf{\mu}=\{\mu_0,\mu_1,\mu_2\}$ and variance covariance matrix $\sigma^2\cal{I}$ where $\cal{I}$ is a $3\times3$ identity matrix; $\mathbf{\phi}=\{\phi_0,\phi_1,\phi_2\}$ follows a multivariate normal distribution with mean $\mathbf{\omega}=\{\omega_0,\omega_1,\omega_2\}$ and the same variance covariance matrix; $a_j$ and $b_k$ are ``effective doses" instead of actual clinical values.
This method omits the interaction effects between two agents.

The authors provided recommendations about selecting priors and methods to calculate ``effective doses". They used the fact that $\frac{K\tilde{\pi}_{11}}{K(1-\tilde{\pi}_{11})} = \frac{exp\{\mu_0\}}{exp\{\omega_0\}}$ to obtain the solutions for $\mu_0$ and $\omega_0$:
$$\mu_0=log(K\tilde{\pi}_{11}), \omega_0=log(K(1-\tilde{\pi}_{11})).$$
They suggested setting $\mu_1=\mu_2=\omega_1=\omega_2=2\sqrt{\sigma^2}$ and selecting $\sigma^2\in [5,10]$.
They set $a_1=b_1=0$, then 
$$a_j=(\mu_1+\omega_1)^{-1}\log(\tilde{OR}_{j.}),$$ $$b_k=(\mu_2+\omega_2)^{-1}\log(\tilde{OR}_{.k})$$
where 
$$\tilde{OR}_{j.}=\exp\{\frac{\tilde{\pi}_{j1}/(1-\tilde{\pi}_{j1})}{\tilde{\pi}_{11}/(1-\tilde{\pi}_{11})}\},$$ $$\tilde{OR}_{.k}=\exp\{\frac{\tilde{\pi}_{1k}/(1-\tilde{\pi}_{1k})}{\tilde{\pi}_{11}/(1-\tilde{\pi}_{11})}\}.$$
Therefore, this design needs inputs of $\pi_{j1}$ and $\pi_{1k}$, $j=1,2,\dots,J, k=1,2,\dots,K$ from the clinicians.

\textbf{Start-up phase}:
this method is a single stage design without a start-up phase.

\textbf{Escalation / de-escalation dose set}:
if the current combination is $(j,k)$, acceptable dose combination set to the next subject $S$ is defined as $(j-1,k)$, $(j+1,k)$, $(j,k)$, $(j,k+1)$, $(j,k-1)$, $(j+1,k-1)$, $(j-1,k+1)$, $(j+1,k+1)$, $(j-1,k-1)$.
As dose combinations $(j+1,k+1)$ and $(j-1,k-1)$ are included, this design allows simultaneous dose escalation or de-escalation of both agents.

\textbf{Trial conduct}:
\renewcommand{\labelenumi}{\alph{enumi}}
\begin{enumerate}
    \item Compute a $95\%$ CI for overall toxicity rate from cumulative data of currently recruited subjects.
    \item If the lower bound of this CI is greater than $\phi$, terminate the trial.
    \item If the lower bound of this CI is less than or equal to $\phi$, use all previous information to obtain posterior mean of $\pi_{jk}$, $j=1,2,\dots,J, k=1,2,\dots,K$.
    \item Select a dose that belongs to set $S$ and is closest to $\phi$. Assign this dose combination to the next patient.
    \item Continue until all $N$ subjects are exhausted.
\end{enumerate}

\textbf{MTD determination}:
Use the outcomes and assignments of all $N$ subjects to derive posterior mean of $\pi_{jk}$, $j=1,2,\dots,J, k=1,2,\dots,K$. If the last subject received dose $(j',k')$, then the dose combination that is among set $S$ of $(j',k')$ and with estimated toxicity closest to $\phi$ will be selected as the MTD.

\subsection[DFCOMB]{Model-based: Logistic model-Based Bayesian dose finding design (DFCOMB)}
\textbf{Dose-toxicity model}: 
this method uses logistic regression to link doses and toxicities of the two agents:
\begin{equation} \label{eq9}
\text{logit}(\pi_{jk})=\beta_0+\beta_1a_j+\beta_2b_k+\beta_3a_jb_k,
\end{equation}
where $a_j$ and $b_k$ are ``effective doses" instead of actual clinical values, $\beta_1>0$, $\beta_2>0$, $\beta_1+\beta_3b_k>0$, $\beta_2+\beta_3a_j>0$ to ensure monotonicity.
The authors define ``effective doses" as $a_j=\log(\frac{p_j}{1-p_j})$, $b_k=\log(\frac{q_k}{1-q_k})$ and recommend a vague normal prior $N(0,1)$ for $\beta_0$ and $\beta_3$, an informative prior exp$\{1\}$ for $\beta_1$ and $\beta_2$.

\textbf{Start-up phase}:
the start-up phase starts from dose $(1,1)$. If no toxicity is observed, escalate the dose along the diagonal until at least one agent reaches maximum dose. If still no toxicity is observed when one agent reaches its maximum dose, we increase the dose of the other agent until both agents reach maximum doses.
The start-up phase ends once the first toxicity is observed and the model-based design starts. 

\textbf{Post start-up escalation / de-escalation dose set}:
if the current combination is $(j,k)$, the dose escalation set is defined as ${\cal A}_E=\{(j+1,k),(j,k+1),(j+1,k-1),(j-1,k+1)\}$, dose de-escalation set is define as ${\cal A}_D=\{(j-1,k),(j,k-1),(j+1,k-1),(j-1,k+1)\}$.
As we only know the partial order of dose toxicity in combinational agents, we do not know whether dose combinations $(j+1,k-1)$ and $(j-1,k+1)$ are more/less toxic than dose combination $(j,k)$. Therefore, the authors included $(j+1,k-1)$ and $(j-1,k+1)$ in both escalation and de-escalation sets. 

\textbf{Post start-up trial conduct}:
in the model-based design part, the escalation and de-escalation rule is the same as in Copula design \citep{yin2009bayesian}. 

\textbf{MTD determination}:
DFCOMB utilizes a different method to identify MTD after the trial is completed. The dose combination that has the largest posterior probability $P(\pi_{jk}\in [\phi-\delta, \phi+\delta])$ and is used to treat at least one cohort will be selected as the MTD. Parameter $\delta$ is the length around the target toxicity probability.

\subsection[gCRM]{Model-based: A Generalized Continual Reassessment Method (gCRM)}
This method is another generalization of the CRM. 

\textbf{Dose-toxicity model}:
it uses proportional odds logistic regression to model the dose-toxicity relationship:
\begin{equation} \label{eq10}
\text{logit}(\pi_{jk})=\alpha_k+\beta a_j,
\end{equation}
where $\alpha_k$ is agent $B$ specific intercept, $k=1,2,\dots,K$; $\beta$ is a common coefficient across models; $a_j$ is the ``effective dose" of agent $A$, $j=1,2,\dots,J$.
For example, if agent $B$ has three dose levels, then gCRM will need three models: $\text{logit}(\pi_{j1})=\alpha_1+\beta a_j$, $\text{logit}(\pi_{j2})=\alpha_2+\beta a_j$, and $\text{logit}(\pi_{j3})=\alpha_3+\beta a_j$. Later these ``sub" models will be aggregated together through a joint prior distribution that forces correlation among $(\alpha_1,\alpha_2,\dots,\alpha_k)$. As observed, gCRM assumes no interaction between two agents as well.

In terms of parameters $\alpha_k$ and $\beta$, their paper assumes that $\beta$ follows a Gamma distribution with mean $\mu_\beta$ and variance $\sigma^2_\beta$, $\alpha_1\sim N(\mu_{\alpha}, \sigma^2_\alpha)$, and defines $\Delta_k=\alpha_k-\alpha_{k-1}\sim N(\delta_k,2\sigma^2_\alpha)$ for $k=2,3,\dots,K$ so the joint distribution of $\mathbf{\alpha}=(\alpha_1,\dots,\alpha_K)^T$ is multivariate normal.
If one assumes that $\text{logit}(\pi_{j1})=E(\alpha_1)+E(\beta)a_j$, $a_j\approx [\text{logit}(\pi_{j1})-\mu_\alpha]/\mu_\beta$ can be obtained. One can approximately obtain $\delta_k=\text{logit}(\pi_{1k}) - \text{logit}(\pi_{1,k-1})$. The authors recommend setting $\mu_\alpha=-8$, $\mu_\beta=1$,  $\sigma^2_\alpha=\sigma^2_\beta =1$. Therefore, with the inputs of $\pi_{j1}$ and $\pi_{1k}$ for $j=1,2,\dots,J, k=1,2,\dots,K$ from clinicians, all parameters can be calculated.

\textbf{Start-up phase}:
this method is a single stage design without a start-up phase.

\textbf{Escalation / de-escalation dose set}:
if the current combination is $(j,k)$, define acceptable dose combination set to the next subject $S$ to be $(j-1,k)$, $(j+1,k)$, $(j,k)$, $(j,k+1)$, $(j,k-1)$, $(j+1,k-1)$, $(j-1,k+1)$, $(j+1,k+1)$, $(j-1,k-1)$.
As dose combination $(j+1,k+1)$ and $(j-1,k-1)$ are included, this design allows simultaneous dose escalation or de-escalation of both agents.

\textbf{Trial conduct}:
\renewcommand{\labelenumi}{\alph{enumi}}
\begin{enumerate}
    \item Treat the first patient at dose $(1,1)$. 
    \item For patient $2,3,\dots, N$, compute $\hat{\pi}_{jk}$ from logit$(\hat{\pi}_{jk})=\hat{\alpha}_k+\hat{\beta}a_j$ where $\hat{\alpha}_k$ and $\hat{\beta}$ are posterior means.
    \item As the posterior distribution of $\pi_{11}$ will be updated constantly, check that whether the stopping rule of $P(\pi_{11}>\phi)>0.95$ has been reached. If yes, then terminate the trial; otherwise assign next subject to the dose in set $S$ where $\hat{\pi}_{jk}$ is closest to $\phi$.
    \item Continue until all $N$ subjects are exhausted. 
\end{enumerate}

\textbf{MTD determination}:
if the last subject received dose $(j',k')$, then the dose combination that is among set $S$ of $(j',k')$ and with estimated toxicity closest to $\phi$ will be selected as the MTD.

\subsection[bCRM]{Model-based: Bootstrap Aggregating Continual Reassessment Method (bCRM)}
Bootstrap aggregating CRM is similar to POCRM as they both use one-dimensional CRM to identify the MTD. 
However, bCRM keeps updating the toxicity ordering of dose combinations rather than pre-specifying them. 

\textbf{Dose-toxicity model}:
in bCRM, it assigns a beta prior to $\pi_{jk}$ and obtain its posterior mean $\bar{\pi}_{jk}$. Then bCRM applies two-dimensional pool-adjacent-violators algorithm (PAVA) \citep{bril1984algorithm} on $\bar{\pi}_{jk}$ to obtain $\tilde{\pi}_{jk}$ to ensure that these estimates satisfy partial ordering. To avoid ties among $\tilde{\pi}_{jk}$, a term $r_{jk}\epsilon$ is added, where $r_{jk}$ is the rank of dose $(j,k)$ and $\epsilon$ is a small positive number. 
The resulted estimates are denoted as $\tilde{\pi}^{\dagger}_{jk}$ and, as a result, one can obtain a new ordering $\cal{O}$. As noted by the authors, such orderings could vary dramatically due to data sparsity. 
Therefore, they bootstrapped $B$ samples of data to obtain corresponding orderings ${\cal O}_b$, and toxicity probability estimates $\hat{\pi}^{\text{b}}_{jk}$,  $b \in 1,2,\dots,B$. The final estimate of $\pi_{jk}$ is
\begin{equation} \label{eq11}
\hat{\pi}^{\text{Bagging}}_{jk}= \sum^{B}_{b=1}P({\cal O}_b|D)\hat{\pi}^{\text{b}}_{jk},
\end{equation}
where D = 
$\begin{bmatrix}
t_1 & \dots & t_n \\
d_1 & \dots & d_n 
\end{bmatrix}$ represents cumulative data up to $n^{th}$ subject, $t_i$ indicates whether subject $i$ experienced DLT or not; $d_i$ indicates dose combination subject $i$ received, $i=1,2,\dots,n$.  

\textbf{Start-up phase}:
the start-up phase is similar to DFCOMB.

\textbf{Post start-up escalation / de-escalation dose set}:
similar to POCRM, bCRM uses one-dimensional CRM in the dose-finding process, therefore, the escalation and de-escalation dose is certain within each ordering.

\textbf{Post start-up trial conduct}:
the trial conduct procedures are similar to DFCOMB.

\textbf{MTD determination}:
After the trial is completed, one could select the combination that has been administered to patients and has the largest posterior probability of falling into the $\varepsilon$-neighbourhood of $\phi$, where $\varepsilon$ is a small positive number.

\subsection[cBOIN]{Model-assisted: Combinational Bayesian optimal interval design (cBOIN)}
Combinational BOIN is a model-assisted design that is generalized from the single agent BOIN design \citep{liu2015bayesian,yuan2016bayesian}.  

\textbf{Dose escalation and de-escalation rule}:
BOIN mainly involves two important parameters $\Delta_L$ and $\Delta_U$ which are lower and upper cut-offs. 
At current dose $j$, the escalation and de-escalation rules are below:
\begin{itemize}
    \item if $\hat{p_j}\in (\phi-\Delta_L,\phi+\Delta_U)$, then the next cohort stays at current dose;
    \item if $\hat{p_j}\leq \phi-\Delta_L$, then the next cohort escalates to dose $j+1$;
    \item if $\hat{p_j}\geq \phi+\Delta_U$, then the next cohort de-escalates to dose $j-1$;
\end{itemize}
where $\hat{p_j}$ is the estimated toxicity probability of dose $j$ in single agent dose-finding and it is simply proportion of patients experiencing toxicities among those who receive dose $j$.

In two-dimensional dose-finding, $\hat{p}_{jk}$ is calculated the same way: $\hat{p}_{jk}=y_{jk}/n_{jk}$.  

An important task of cBOIN is to determine $\Delta_L$ and $\Delta_U$. Through minimizing the probability of incorrect movement given data at current dose,
$$\Delta_L=\phi-\frac{\log\{\frac{1-\phi_1}{1-\phi}\}}{\log\{\frac{\phi(1-\phi_1)}{\phi_1(1-\phi)}\}},
\Delta_U=\frac{\log\{\frac{1-\phi}{1-\phi_2}\}}{\log\{\frac{\phi_2(1-\phi)}{\phi(1-\phi_2)}\}}-\phi.$$
The authors suggested using $\phi_1=0.6\phi$ and $\phi_2=1.4\phi$ through their simulation calibration.

\textbf{Start-up phase}:
this method does not have a start-up phase.

\textbf{Escalation / de-escalation dose set}:
admissible dose escalation set is defined as ${\cal A}_E=\{(j+1,k),(j,k+1)\}$, admissible de-escalation set is define as ${\cal A}_D=\{(j-1,k),(j,k-1)\}$.

\textbf{Trial conduct}:
\renewcommand{\labelenumi}{\alph{enumi}}
\begin{enumerate}
    \item Treat the first cohort at dose $(1,1)$.
    \item Suppose that current dose is combination $(j,k)$. If $\hat{p}_{jk}\leq \phi-\Delta_L$, escalate to the dose combination that belongs to ${\cal A}_E$ and has the largest $P[p_{j\prime k\prime}\in (\phi-\Delta_L,\phi+\Delta_U)|y_{j\prime k\prime}]$.
    \item If $\hat{p}_{jk}\geq \phi+\Delta_U$, de-escalate to the dose combination that belongs to ${\cal A}_D$ and has the largest $P[p_{j\prime k\prime}\in (\phi-\Delta_L,\phi+\Delta_U)|y_{j\prime k\prime}]$.
    \item Otherwise if $\phi-\Delta_L<\hat{p}_{jk}<\phi+\Delta_U$, stay at current dose.
    \item Dose combinations with $P(p_{jk}>\phi|y_{jk})\geq \lambda$ will be permanently excluded, where $\lambda$ is pre-specified threshold probability. If dose combination $(1,1)$ satisfies this stopping rule, the trial will be terminated early.
    \item Continue until all $N$ subjects are exhausted.
\end{enumerate}

\textbf{MTD determination}:
after the trial is completed, isotonic regression will be used on $\hat{p}_{jk}$ to obtain estimator $\tilde{p}_{jk}$ so that they satisfy monotonic dose-toxicity within one agent when fixing the other agent's dose. The MTD is the dose combination with $\tilde{p}_{jk}$ closest to $\phi$.

\subsection[cKeyboard]{Model-assisted: Combinational Keyboard design (cKeyboard)}
Similar to combinational BOIN, combinational Keyboard is a model-assisted design as well.
Combinational Keyboard design starts with specifying a target toxicity interval ${\cal J}_{target}=(\phi-\varepsilon_1,\phi+\varepsilon_2)$, where $\varepsilon_1$ and $\varepsilon_2$ are tolerable deviations from $\phi$. This interval ${\cal J}_{target}$ is called target key.
Then a series of equal-width keys are identified along both sides of the target key.

\textbf{Dose escalation and de-escalation rule}:
in the setting of the single agent design, the escalation and de-escalation rules are straightforward. Define ${\cal J}_{\text{max}}$ to be the strongest key based on the posterior distribution of current dose $j$,
\begin{itemize}
    \item if ${\cal J}_{\text{max}}\prec {\cal J}_{target}$, then next cohort escalates to dose $j+1$;
    \item if ${\cal J}_{\text{max}}\equiv {\cal J}_{target}$, then next cohort stays at current dose;
    \item if ${\cal J}_{\text{max}}\succ {\cal J}_{target}$, then next cohort de-escalates to dose $j-1$;
\end{itemize}

To address two-dimensional dose-finding, the authors define five strategies of admissible escalate and de-escalate sets. After simulations, the strategy whose admissible escalation and de-escalation sets are the same with combinational BOIN design is recommended.

\textbf{Start-up phase}:
this method does not have a start-up phase.

\textbf{Escalation / de-escalation dose set}:
several dose assignment algorithms have been proposed for Keyboard design and the authors recommend to define
admissible dose escalation set to be ${\cal A}_E=\{(j+1,k),(j,k+1)\}$, admissible de-escalation set to be ${\cal A}_D=\{(j-1,k),(j,k-1)\}$.

\textbf{Trial conduct}:
\renewcommand{\labelenumi}{\alph{enumi}}
\begin{enumerate}
    \item Treat the first cohort at dose $(1,1)$.
    \item Suppose that current dose is combination $(j,k)$. If ${\cal J}_{\text{max}}\prec {\cal J}_{target}$, escalate to dose combination that belongs to ${\cal A}_E$ and has the largest $P[p_{j\prime k\prime}\in {\cal J}_{target}|(n_{jk},y_{jk)}]$.
    \item If ${\cal J}_{\text{max}}\succ {\cal J}_{target}$, de-escalate to dose combination that belongs to ${\cal A}_D$ and has the largest $P[p_{j\prime k\prime}\in {\cal J}_{target}|(n_{jk},y_{jk})]$.
    \item Otherwise if ${\cal J}_{\text{max}}\equiv {\cal J}_{target}$, stay at current dose.
    \item Dose combinations with $P(p_{jk}>\phi|y_{jk})\geq \lambda$ will be permanently excluded, where $\lambda$ is pre-specified threshold probability. If dose combination $(1,1)$ satisfies this stopping rule, the trial will be terminated early.
    \item Continue until all $N$ subjects are exhausted.
\end{enumerate}

\textbf{MTD determination}:
after the trial is completed, isotonic regression will be used to identify the MTD.

\section{Simulation Studies}\label{sec3}
 
In the simulation studies, our goal is to identify single MTD of two combined agents. 
Simulation settings are borrowed from previous studies \citep{riviere2015competing, hirakawa2013dose} and shown in Table~\ref{setting}. 
The target toxicity probability is 0.3. 
All designs started with the lowest dose combination. 
Cohort size was set to be 3 for all designs that use cohorts as dose assignment unit, unless otherwise specified. 
2000 simulation runs were generated for each scenario.

\subsection{Simulation scenarios}

A total of 15 scenarios are displayed in Table~\ref{setting}. In the first 10 scenarios, agent $A$ has 5 dose levels and agent $B$ has 3. In scenarios 11 to 15, both agents have 4 dose levels. Target toxicity rate 0.3 is bolded. Among the first ten $5 \times 3$ matrices: scenario 1 contains multiple MTD locations that are in the middle of matrix and diagonally connected; scenarios 2 and 4 represent over-toxic situations while scenario 4 is more extreme; scenarios 3 and 5 represent over-conservative situations while scenario 5 is more extreme; scenarios 6 and 7 contain multiple MTD locations but those locations are more scattered; scenarios 8, 9, and 10 contain single MTD at different locations. Among the last five $4 \times 4$ square matrices: scenario 11 contains multiple MTD locations that are in the middle of matrix and diagonally connected; scenarios 12 and 13 contain multiple but more scattered MTD locations; scenario 14 contains two MTD locations that are at the bottom left and top right; scenario 15 has single MTD location.

\subsection{Evaluation metrics}

Four evaluation metrics are used: (1) correct MTD selection $S_{C}$, defined as proportion of simulation runs that correctly identified the MTD among all 2000 simulations; (2) over-toxic MTD selection $S_{OT}$, defined as the proportion of simulation runs that identified over-toxic doses as MTD among all 2000 simulations; (3) correct patient assignment $A_{C}$, defined as the average proportion of patients that were assigned to the MTD during the trial across all 2000 simulations; (4) over-toxic patient assignment $A_{OT}$, defined as the average proportion of patients that were assigned to over-toxic doses during the trial across all 2000 simulations. 
Metrics $S_{C}$ and $S_{OT}$ will be used to evaluate the performance of designs in terms of MTD selection. 
The larger the $S_{C}$ is, the more accurate the design is in selecting the correct MTD. The larger the $S_{OT}$ is, the more aggressive the design is in selecting MTD. 
Metrics $A_{C}$ and $A_{OT}$ will be used to evaluate the characteristics of designs during trial conduct. The larger the $A_{C}$ is, the more accurate patient assignment is during the trial. The larger the $A_{OT}$ is, the more aggressive the design is in dose escalation during the trial. 
Ideally, a design should show relatively large $S_{C}$ and $A_{C}$ but small $S_{OT}$ and $A_{OT}$.

\subsection{Design specifications}

For I2D, we implemented published R codes \citep{ezzalfani2019design}.
We set cohort size of start-up phase to be 1 based on suggestions from simulation studies when target toxicity rate is 0.3 \citep{ivanova2003improved, wang2005two}, and interaction to be 0 so that it is consistent with the paper's focus. 
The prior of parameters $(\alpha,\beta)$ is the product of two independent exponential distributions with mean 1, which is the same as the one used in the I2D study \citep{wang2005two}.

For Copula, the website (\url{http://www.blackwellpublishing.com/rss}) where simulation programs were originally published is not accessible now, so we used the executable file on the website \url{https://odin.mdacc.tmc.edu/~yyuan/index_code.html}. 
The executable file used escalation and de-escalation probability boundaries fixed at 0.8 and 0.45, respectively. 

Hierarchy was implemented via R codes from website \url{http://www-personal.umich.edu/~tombraun/software.html}. We set $\sigma^2$ to be 10 based on the suggestion in the paper \citep{braun2010hierarchical}. Together with other recommendations from the authors, we could obtain priors for all involved parameters. Details are described in Section ~\ref{hierarchy}.

POCRM was implemented via R package \texttt{pocrm}. We utilized six possible partial ordering as suggested \citep{wages2013pocrm, wages2013specifications}: across rows, across columns, up diagonals, down diagonals, up-down diagonals, and down-up diagonals. 
As we do not have information about which partial ordering is more likely, the prior probabilities of all 6 partial ordering were set to be equal. 
The skeleton required by the program was obtained using \texttt{getprior} function in package \texttt{dfcrm} from algorithm of Lee and Cheung \citep{lee2009model} as suggested \citep{wages2013pocrm}. 
For the start-up phase, we used the “zoning” method as suggested \citep{wages2011continual, wages2011dose}. 

Design cBOIN was implemented via R package \texttt{BOIN}. For cBOIN, the interval boundaries were set to be 0.18 and 0.42 as suggested \citep{lin2017bayesian}.

Design cKeyboard was implemented via R package \texttt{Keyboard}.

DFCOMB was implemented via R package \texttt{dfcomb}.
As recommended by the authors, a vague normal prior $N(0,1)$ for $\beta_0$ and $\beta_3$, an informative prior exp$\{1\}$ for $\beta_1$ and $\beta_2$ \citep{riviere2014bayesian}.
For DFCOMB, we set the target toxicity boundaries as 0.18 and 0.42 to be consistent with cBOIN. 
As one of our reviewers suggested, we also tried setting these toxicity boundaries as 0.25 and 0.35.

Design gCRM was implemented via R codes from website \url{http://www-personal.umich.edu/~tombraun/software.html}. As the authors suggested, we used a Gamma prior with mean 1 and variance 1 for $\beta$, normal prior with mean -8 and variance 1 for $\alpha_1$, normal prior with mean $\text{logit}(\pi_{1k}) - \text{logit}(\pi_{1,k-1})$ and variance 2 for $\delta_k$, where $k=2,3,\dots,K$ \citep{braun2013generalized}.

Design bCRM was implemented via R codes from the authors. Its skeleton setting is the same as POCRM. 

For most model-based designs, there are several design parameters involved. Some of these parameters have recommended specifications provided by the authors, for example, interval boundaries in design cBOIN are recommended to set as $0.6\phi$ and $1.4\phi$ where $\phi$ is the target toxicity probability \citep{lin2017bayesian}. However, some design parameters lack authors' suggested specifications and their influences to design performances are not clear. Therefore, we list such design parameters and corresponding designs in Table~\ref{alter} where column ``Main setting" and column ``Alternative setting" contain parameter specifications used in our simulation studies. 


For all designs and scenarios, the maximum sample size was set to be 60. This number is widely used in other studies \citep{yin2009bayesian, braun2013generalized,lin2016bootstrap,lin2017bayesian,pan2020keyboard,riviere2015competing}. To verify that our conclusions are still valid with a different sample size, we repeated all simulations with a maximum sample size of 30.

\section{Results}\label{sec4}

\subsection{When maximum sample size is 60}

Table~\ref{mtd}, Table~\ref{toxicmtd}, Table~\ref{patMTD}, and Table~\ref{patToxic} display design performances of $S_{C}$, $S_{OT}$, $A_{C}$, and $A_{OT}$, respectively. 
In these tables, I2D with parameter $p_j$ and $q_k$ specification in column ``Main setting" of Table~\ref{alter} is denoted as ``I2D"; I2D with parameter $p_j$ and $q_k$ specification in column ``Alternative setting" is denoted as ``I2D.pq". 
Copula with parameter $p_j$ and $q_k$ specification in column ``Main setting" is denoted as ``Copula"; Copula with parameter $p_j$ and $q_k$ specification in column ``Alternative setting" is denoted as ``Copula.pq". 
Hierarchy with $\pi_{1k}$ and $\pi_{j1}$ specification in column ``Main setting" is denoted as ``Hierarchy"; Hierarchy with $\pi_{1k}$ and $\pi_{j1}$ specification in column ``Alternative setting" is denoted as ``Hierarchy.pi".
POCRM with skeleton specification in column ``Main setting" is denoted as ``POCRM"; POCRM with skeleton specification in column ``Alternative setting" is denoted as ``POCRM.skeleton". 
DFCOMB with $p_j$ and $q_k$, and escalation/de-escalation probability cutoff specifications in column ``Main setting" is denoted as ``DFCOMB"; DFCOMB with parameter $p_j$ and $q_k$ specification in column ``Alternative setting" is denoted as ``DFCOMB.pq"; DFCOMB with escalation/de-escalation probability cutoff specification in column ``Alternative setting" is denoted as ``DFCOMB.cut"; DFCOMB with $p_j$ and $q_k$, and escalation/de-escalation probability cutoff specifications in column ``Main setting", but with toxicity boundaries being 0.25 and 0.35 as suggested by one of our reviewers, is denoted as ``DFCOMB.sensitive". 
Design gCRM with $\pi_{1k}$ and $\pi_{j1}$ specification in column ``Main setting" is denoted as ``gCRM"; gCRM with $\pi_{1k}$ and $\pi_{j1}$ specification in column ``Alternative setting" is denoted as ``gCRM.pi".
Design bCRM with skeleton and escalation/de-escalation probability cutoff specifications in column ``Main setting" is denoted as ``bCRM"; bCRM with skeleton specification in column ``Alternative setting" is denoted as ``bCRM.skeleton"; bCRM with escalation/de-escalation probability cutoff specification in column ``Alternative setting" is denoted as ``bCRM.cut". 
We marked designs metrics that are ``outstandingly" poor as red, and those that are poor, but not ``outstanding" from the others as magenta.


\textit{I2D} shows unstable performances in MTD identification in most simulation scenarios. 
While under extreme conditions like scenario 4 and 5, its $S_{C}$ is among the best, under scenarios like scenarios 1, 3, 8, 9, and 15, its $S_{C}$ is among the worst. 
With alternative toxicity profile of individual agents ($p_j$ and $q_k$), some scenarios had improved performances while some had worse, but its overall characteristics remain the same. 
Part of the reason of the unstable performance is that, I2D always starts its model-based part with agent $B$'s lowest dose no matter what happens in the start-up phase. 
This way, the starting dose combination of model-based part could be very far from the location of true MTDs, which makes I2D harder to identify them. 
Overall I2D is not an aggressive design as its $S_{OT}$ and $A_{OT}$ are relatively small under most scenarios.

\textit{Copula} performed poorly in MTD identification as its $S_{C}$ and $S_{OT}$ are among the worst in several scenarios. 
This indicates that Copula is quite aggressive as it is more likely to select higher dose combinations as the MTD. 
With alternative toxicity profile of individual agents ($p_j$ and $q_k$), Copula's performances fluctuated in several scenarios, but overall poor performances and aggressiveness were still observed. 
One potential reason of the unsatisfactory performances is the limited flexibility on changing parameters like escalation and de-escalation probability cutoffs in the executable file. 
Therefore, it is reasonable to argue that the default values (0.8 and 0.45) are not optimal for some scenarios. 
On the other hand, as we do not know what the true dose toxicity matrix looks like in real life, obtaining a uniform parameter set to achieve best performances under all scenarios through simulation calibration is not feasible.

\textit{Hierarchy} is quite aggressive overall in trial conduct as it has the worst $A_{OT}$ under several scenarios. 
We can observe that incorrect $\pi_{j1}$ and $\pi_{1k}$ performed much worse than using correct ones in most scenarios. 
A possible reason for the aggressiveness is that, unlike most other designs, Hierarchy allows simultaneous dose escalation of both agents during trial conduct. 
Another aspect we should emphasize is that despite a high proportion of patients assigned to over-toxic doses, Hierarchy did not outperform other designs in terms of $S_{C}$. 
Some features of Hierarchy like omitting the interaction effect between agents and no start-up phase may contribute to this poor performance as well.

\textit{POCRM} shows satisfactory characteristics across all scenarios except low $S_{C}$ in scenario 5. 
Then we found that the alternative skeleton setting in Table~\ref{alter} improved $S_{C}$ from 0.54 to 0.73 in scenario 5. 
However, the alternative skeleton setting led to much worse performance metrics in scenario 15.
Another finding is that POCRM performed well under scenarios (e.g., scenario 9) when the underlying true toxicity orderings of dose combinations are not among any of the six orderings we used.
Such results ``validate'' the idea of POCRM: in practice we do not need to specify the correct toxicity ordering in POCRM, providing orderings close to the correct one is sufficient.

\textit{DFCOMB} performed poorly in terms of $S_{C}$ under several scenarios. 
But overall DFCOMB is not an aggressive design as its $S_{OT}$ and $A_{OT}$ are relatively small under most scenarios. 
With alternative escalation and de-escalation probability cutoffs, some scenarios had improved $S_{C}$ (e.g. scenarios 9 and 10) while some had worse $S_{C}$ (e.g. scenarios 6 and 8). In addition, we observed worse $A_{OT}$ and better $S_{OT}$ with alternative specification (0.6 and 0.6 as escalation and de-escalation probability cutoffs). Worse $A_{OT}$ is expected as the alternative cutoff pairs make DFCOMB easier for dose escalation but more difficult for de-escalation. With more patients assigned to over-toxic doses, the estimation of dose toxicity probabilities were more accurate, which leads to better $S_{OT}$.
With alternative toxicity profile of individual agents ($p_j$ and $q_k$), a few scenarios showed obvious impact: $S_{C}$ in scenario 1 was improved from 0.54 to 0.69, $S_{C}$ in scenario 11 was improved from 0.44 to 0.57, $S_{C}$ in scenario 14 was improved from 0.18 to 0.35, but $S_{C}$ in scenario 10 and 15 were dramatically reduced from 0.47 to 0.14, and from 0.67 to 0.45, respectively. 
With the ``sensitivity run'' of target toxicity boundaries suggested by one of our reviewers, we observed slightly worse $S_{OT}$ in some scenarios.
Such results imply that the optimal design parameters are scenario-dependent. Therefore, it is not feasible for us to calibrate the parameters through simulations in real life clinical trials.

Design \textit{gCRM} performed well in most scenarios. 
Comparing results using correct $\pi_{j1}$ and $\pi_{1k}$ with incorrect ones, we observed that using incorrect inputs leads to worse operating characteristics under some scenarios, and similar performances under the other ones. 
Interestingly, although gCRM allows simultaneous dose escalation of both agents, it did not show much aggressiveness as its $S_{OT}$ and $A_{OT}$ are not among the largest under most scenarios. 
One possible reason could be that gCRM uses single patient as unit, instead of cohorts during trial conduct. 
Therefore, every time when an over-toxic dose combination is assigned, only one patient instead of a cohort of several patients will receive it. 
From this perspective, gCRM could be viewed as more ``flexible'' in the dose-finding process and such flexibility may dilute the aggressiveness.

Design \textit{bCRM} is another one whose performances were unstable across different scenarios. 
Its $S_{C}$ in scenario 5 and $S_{OT}$ in scenario 4 are among the worst. Metrics in other scenarios are acceptable. 
Similar to DFCOMB, we observed worse $A_{OT}$ with alternative escalation and de-escalation probability cutoffs. However, $S_{OT}$ was not improved. 
Alternative skeleton setting has influences to bCRM as well as it improved the $S_{C}$ in scenario 5 and $S_{OT}$ in scenario 4, but worsened $S_{C}$ in scenario 10 and $S_{OT}$ in scenario 9 and 10. 
Therefore, unstable performances remains an issue even using alternative skeleton setting or escalation and de-escalation probability cutoffs.

Both designs \textit{cBOIN} and \textit{cKeyboard} performed well across all scenarios. They may not always be the top performers, but their operating characteristics are never among the worst. 
This is especially important in real life clinical trials, as we do not know which scenario could be the truth. Therefore, cBOIN and cKeyboard are able to guarantee satisfactory performances in practice. 
 
\subsection{When maximum sample size is 30}
 
Results using 30 as the maximum sample size are shown in Supplementary Materials Table 5 to Table 12.
All the findings discussed above are observed when maximum sample size is reduced from 60 to 30. Additionally, we found that in Supplementary Materials Table 5, POCRM with alternative skeleton had unsatisfactory $S_C$ in scenario 15 given maximum sample size of 30. This indicates that parameter calibration is not feasible for POCRM as well.

\section{Discussion}\label{sec5}
Despite recent advances in novel statistical designs for combinational agents, we found that they are seldomly cited and used in ongoing clinical trials.
In dose-finding studies for ``combinational agents'', the investigators often conducted dose-finding for one agent, while the second agent remained fixed.

Riviere et al. \citep{riviere2015designs} reviewed 543  clinical trial papers published between 2011 and 2013 that investigated combinational agents. Among these papers, 162 had at least two agents dose-escalated and the rest (381) had only one agent dose-escalated with the others fixed. 
Only one out of 543 papers used a design that was ideal for combinational agents.
On the website \url{https://clinicaltrials.gov/}, we found 591 phase I/early phase I intervention studies in the U.S. for combinational agents with trial results, with primary completion dates after 1/1/2010; however, the nine designs we evaluate here were only cited in less than 5 trial papers. 
While these 591 trials include trials that have yet to be published, it would appear that optimal designs for combination therapies are underutilized.
The discrepancy between low acceptance of novel designs in clinical practice and the endeavor of promoting better designs should be reconciled.

There are several barriers to implementing more optimal designs for clinical trials exploring combinational agents.
First, there is no practical guidance in terms of design selection to the investigators. 
Second, model-based designs are not easily understood, and are relatively complicated in implementation as they usually require robust assumptions, parameter calibration, and ongoing statistical support to update toxicity probability estimation. 
In addition, the start-up phases of various model-based designs could be quite different from each other. Some designs' start-up phases could even largely influence their operating characteristics. Such complexity places another layer of barrier to the broader usage of model-based designs.  
Motivated by these existing hurdles, our simulation study aims to provide practical recommendations to investigators in designing phase I clinical trials and explore the impact of different design parameters in running model-based designs.

From our simulation results, we observed considerable performance fluctuations for several model-based designs in different scenarios. 
Such unstable performances may be due to assumptions of their specific parametric dose-toxicity relationships. When the assumed relationship fits the true scenario, those designs may result in favorable performances, and vice versa. 
Overall, designs POCRM, gCRM, cBOIN, and cKeyboard perform better than the others regarding our evaluation metrics and we recommend them in future combinational dose-finding studies.

From practical perspective, we would like to promote broader usage of cBOIN and cKeyboard for combination trials. The reasons are multifocal. 
First, cBOIN and cKeyboard have guaranteed stable operating characteristics in all scenarios. This feature is crucial in practice without knowing the truth. 
Second, cBOIN and cKeyboard are convenient as they require neither parameter calibration nor the agents' prior information. 
Finally, cBOIN and cKeyboard are much easier to implement as they are able to provide a dose escalation/de-escalation table before trial conduct (similar to the conventional 3+3 design). 
This feature is ideal for investigators who prefer the 3+3 design over model-based designs due to simplicity, even though 3+3 designs have lower accuracy in MTD identification and more exposure to patients to subtherapeutic doses \citep{simon1997accelerated,reiner1999operating}. 
Our findings are quite consistent with a recently published study from the ASA biopharmaceutical working group \citep{liu2022accuracy} whose aim is to evaluate the accuracy and safety among various Phase I designs for combinational agents. Under the situation of finding one MTD, the ASA working group paper also concludes that combinational BOIN is more attractive than algorithm-based and model-based designs in planning phase I clinical trials for combinational agents.

In addition to evaluating the nine designs, we explored the impact of four design parameters that are commonly encountered in model-based designs and do not have researchers' recommendations: monotherapy toxicity profiles, skeleton settings, dose escalation/de-escalation probability cutoffs, and prior guesses of dose combinations' toxicity probabilities. 
After re-running designs using alternative parameter settings, we found that dose escalation/de-escalation probability cutoffs have negligible impact on design operating characteristics in all scenarios. 
But all other parameters showed impacts on design performances. 
When monotherapy toxicity profiles are available, their impact on design performances is not a big concern.  
However, for dose combinations' toxicity probabilities, it is often difficult to provide accurate guesses. Similarly, it is almost impossible to calibrate the skeleton setting through simulation in practice without knowing the true toxicity profile of dose combinations.

Lastly, we repeated all simulations with a different maximum sample size and observed that almost all findings are consistent.

Our simulation study also has limitations. Some designs, such as gCRM, fix their cohort size to be one, making the performance comparison less fair with other designs that have the flexibility of changing cohort sizes.

Our hope is that this paper will contribute to appropriate and responsible study design utilization for phase I trials with combinational agents. Therefore, heightened awareness of these new designs can only deliver improved results.

\bibliographystyle{apalike}
\bibliography{ref}%

\begin{thebibliography}{}

\bibitem[Babb et~al., 1998]{babb1998cancer}
Babb, J., Rogatko, A., and Zacks, S. (1998).
\newblock Cancer phase i clinical trials: efficient dose escalation with
  overdose control.
\newblock {\em Statistics in medicine}, 17(10):1103--1120.

\bibitem[Braun and Jia, 2013]{braun2013generalized}
Braun, T.~M. and Jia, N. (2013).
\newblock A generalized continual reassessment method for two-agent phase i
  trials.
\newblock {\em Statistics in Biopharmaceutical Research}, 5(2):105--115.

\bibitem[Braun and Wang, 2010]{braun2010hierarchical}
Braun, T.~M. and Wang, S. (2010).
\newblock A hierarchical bayesian design for phase i trials of novel
  combinations of cancer therapeutic agents.
\newblock {\em Biometrics}, 66(3):805--812.

\bibitem[Bril et~al., 1984]{bril1984algorithm}
Bril, G., Dykstra, R., Pillers, C., and Robertson, T. (1984).
\newblock Algorithm as 206: isotonic regression in two independent variables.
\newblock {\em Journal of the Royal Statistical Society. Series C (Applied
  Statistics)}, 33(3):352--357.

\bibitem[Conaway et~al., 2004]{conaway2004designs}
Conaway, M.~R., Dunbar, S., and Peddada, S.~D. (2004).
\newblock Designs for single-or multiple-agent phase i trials.
\newblock {\em Biometrics}, 60(3):661--669.

\bibitem[Diniz et~al., 2018]{diniz2018bayesian}
Diniz, M.~A., Kim, S., and Tighiouart, M. (2018).
\newblock A bayesian adaptive design in cancer phase i trials using dose
  combinations in the presence of a baseline covariate.
\newblock {\em Journal of probability and statistics}, 2018.

\bibitem[Ezzalfani, 2019]{ezzalfani2019design}
Ezzalfani, M. (2019).
\newblock How to design a dose-finding study on combined agents: Choice of
  design and development of r functions.
\newblock {\em Plos one}, 14(11):e0224940.

\bibitem[Harrington et~al., 2013]{harrington2013adaptive}
Harrington, J.~A., Wheeler, G.~M., Sweeting, M.~J., Mander, A.~P., and Jodrell,
  D.~I. (2013).
\newblock Adaptive designs for dual-agent phase i dose-escalation studies.
\newblock {\em Nature Reviews Clinical Oncology}, 10(5):277.

\bibitem[Hirakawa et~al., 2013]{hirakawa2013dose}
Hirakawa, A., Hamada, C., and Matsui, S. (2013).
\newblock A dose-finding approach based on shrunken predictive probability for
  combinations of two agents in phase i trials.
\newblock {\em Statistics in medicine}, 32(26):4515--4525.

\bibitem[Hirakawa et~al., 2015]{hirakawa2015comparative}
Hirakawa, A., Wages, N.~A., Sato, H., and Matsui, S. (2015).
\newblock A comparative study of adaptive dose-finding designs for phase i
  oncology trials of combination therapies.
\newblock {\em Statistics in medicine}, 34(24):3194--3213.

\bibitem[Ivanova and Kim, 2009]{ivanova2009dose}
Ivanova, A. and Kim, S.~H. (2009).
\newblock Dose finding for continuous and ordinal outcomes with a monotone
  objective function: a unified approach.
\newblock {\em Biometrics}, 65(1):307--315.

\bibitem[Ivanova et~al., 2003]{ivanova2003improved}
Ivanova, A., Montazer-Haghighi, A., Mohanty, S.~G., and D.~Durham, S. (2003).
\newblock Improved up-and-down designs for phase i trials.
\newblock {\em Statistics in medicine}, 22(1):69--82.

\bibitem[Ivanova and Wang, 2004]{ivanova2004non}
Ivanova, A. and Wang, K. (2004).
\newblock A non-parametric approach to the design and analysis of
  two-dimensional dose-finding trials.
\newblock {\em Statistics in Medicine}, 23(12):1861--1870.

\bibitem[Korn and Simon, 1993]{korn1993using}
Korn, E.~L. and Simon, R. (1993).
\newblock Using the tolerable-dose diagram in the design of phase i combination
  chemotherapy trials.
\newblock {\em Journal of Clinical Oncology}, 11(4):794--801.

\bibitem[Lee and Fan, 2012]{lee2012two}
Lee, B.~L. and Fan, S.~K. (2012).
\newblock A two-dimensional search algorithm for dose-finding trials of two
  agents.
\newblock {\em Journal of biopharmaceutical statistics}, 22(4):802--818.

\bibitem[Lee and Cheung, 2009]{lee2009model}
Lee, S.~M. and Cheung, Y.~K. (2009).
\newblock Model calibration in the continual reassessment method.
\newblock {\em Clinical Trials}, 6(3):227--238.

\bibitem[Lin and Yin, 2017]{lin2017bayesian}
Lin, R. and Yin, G. (2017).
\newblock Bayesian optimal interval design for dose finding in drug-combination
  trials.
\newblock {\em Statistical methods in medical research}, 26(5):2155--2167.

\bibitem[Lin et~al., 2016]{lin2016bootstrap}
Lin, R., Yin, G., et~al. (2016).
\newblock Bootstrap aggregating continual reassessment method for dose finding
  in drug-combination trials.
\newblock {\em The Annals of Applied Statistics}, 10(4):2349--2376.

\bibitem[Liu et~al., 2022]{liu2022accuracy}
Liu, R., Yuan, Y., Sen, S., Yang, X., Jiang, Q., Li, X., Lu, C., G{\"o}neng,
  M., Tian, H., Zhou, H., et~al. (2022).
\newblock Accuracy and safety of novel designs for phase i drug-combination
  oncology trials.
\newblock {\em Statistics in Biopharmaceutical Research},
  (just-accepted):1--19.

\bibitem[Liu and Ning, 2013]{liu2013bayesian}
Liu, S. and Ning, J. (2013).
\newblock A bayesian dose-finding design for drug combination trials with
  delayed toxicities.
\newblock {\em Bayesian analysis}, 8(3):703.

\bibitem[Liu and Yuan, 2015]{liu2015bayesian}
Liu, S. and Yuan, Y. (2015).
\newblock Bayesian optimal interval designs for phase i clinical trials.
\newblock {\em Journal of the Royal Statistical Society: Series C: Applied
  Statistics}, pages 507--523.

\bibitem[Love et~al., 2017]{love2017embracing}
Love, S.~B., Brown, S., Weir, C.~J., Harbron, C., Yap, C., Gaschler-Markefski,
  B., Matcham, J., Caffrey, L., McKevitt, C., Clive, S., et~al. (2017).
\newblock Embracing model-based designs for dose-finding trials.
\newblock {\em British journal of cancer}, 117(3):332--339.

\bibitem[O'Quigley et~al., 1990]{o1990continual}
O'Quigley, J., Pepe, M., and Fisher, L. (1990).
\newblock Continual reassessment method: a practical design for phase 1
  clinical trials in cancer.
\newblock {\em Biometrics}, pages 33--48.

\bibitem[Pan et~al., 2020]{pan2020keyboard}
Pan, H., Lin, R., Zhou, Y., and Yuan, Y. (2020).
\newblock Keyboard design for phase i drug-combination trials.
\newblock {\em Contemporary Clinical Trials}, page 105972.

\bibitem[Reiner et~al., 1999]{reiner1999operating}
Reiner, E., Paoletti, X., and O'Quigley, J. (1999).
\newblock Operating characteristics of the standard phase i clinical trial
  design.
\newblock {\em Computational Statistics \& Data Analysis}, 30(3):303--315.

\bibitem[Riviere et~al., 2015a]{riviere2015competing}
Riviere, M.-K., Dubois, F., and Zohar, S. (2015a).
\newblock Competing designs for drug combination in phase i dose-finding
  clinical trials.
\newblock {\em Statistics in medicine}, 34(1):1--12.

\bibitem[Riviere et~al., 2015b]{riviere2015designs}
Riviere, M.-K., Le~Tourneau, C., Paoletti, X., Dubois, F., and Zohar, S.
  (2015b).
\newblock Designs of drug-combination phase i trials in oncology: a systematic
  review of the literature.
\newblock {\em Annals of Oncology}, 26(4):669--674.

\bibitem[Riviere et~al., 2014]{riviere2014bayesian}
Riviere, M.-K., Yuan, Y., Dubois, F., and Zohar, S. (2014).
\newblock A bayesian dose-finding design for drug combination clinical trials
  based on the logistic model.
\newblock {\em Pharmaceutical statistics}, 13(4):247--257.

\bibitem[Simon et~al., 1997]{simon1997accelerated}
Simon, R., Rubinstein, L., Arbuck, S.~G., Christian, M.~C., Freidlin, B., and
  Collins, J. (1997).
\newblock Accelerated titration designs for phase i clinical trials in
  oncology.
\newblock {\em Journal of the National Cancer Institute}, 89(15):1138--1147.

\bibitem[Thall et~al., 2003]{thall2003dose}
Thall, P.~F., Millikan, R.~E., Mueller, P., and Lee, S.-J. (2003).
\newblock Dose-finding with two agents in phase i oncology trials.
\newblock {\em Biometrics}, 59(3):487--496.

\bibitem[Tighiouart et~al., 2017]{tighiouart2017bayesian}
Tighiouart, M., Li, Q., and Rogatko, A. (2017).
\newblock A bayesian adaptive design for estimating the maximum tolerated dose
  curve using drug combinations in cancer phase i clinical trials.
\newblock {\em Statistics in medicine}, 36(2):280--290.

\bibitem[Wages and Conaway, 2013]{wages2013specifications}
Wages, N.~A. and Conaway, M.~R. (2013).
\newblock Specifications of a continual reassessment method design for phase i
  trials of combined drugs.
\newblock {\em Pharmaceutical statistics}, 12(4):217--224.

\bibitem[Wages et~al., 2011a]{wages2011continual}
Wages, N.~A., Conaway, M.~R., and O'Quigley, J. (2011a).
\newblock Continual reassessment method for partial ordering.
\newblock {\em Biometrics}, 67(4):1555--1563.

\bibitem[Wages et~al., 2011b]{wages2011dose}
Wages, N.~A., Conaway, M.~R., and O'Quigley, J. (2011b).
\newblock Dose-finding design for multi-drug combinations.
\newblock {\em Clinical Trials}, 8(4):380--389.

\bibitem[Wages and Varhegyi, 2013]{wages2013pocrm}
Wages, N.~A. and Varhegyi, N. (2013).
\newblock pocrm: an r-package for phase i trials of combinations of agents.
\newblock {\em Computer methods and programs in biomedicine}, 112(1):211--218.

\bibitem[Wang and Ivanova, 2005]{wang2005two}
Wang, K. and Ivanova, A. (2005).
\newblock Two-dimensional dose finding in discrete dose space.
\newblock {\em Biometrics}, 61(1):217--222.

\bibitem[Yan et~al., 2017]{yan2017keyboard}
Yan, F., Mandrekar, S.~J., and Yuan, Y. (2017).
\newblock Keyboard: a novel bayesian toxicity probability interval design for
  phase i clinical trials.
\newblock {\em Clinical Cancer Research}, 23(15):3994--4003.

\bibitem[Yin and Yuan, 2009a]{yin2009bayesian}
Yin, G. and Yuan, Y. (2009a).
\newblock Bayesian dose finding in oncology for drug combinations by copula
  regression.
\newblock {\em Journal of the Royal Statistical Society: Series C (Applied
  Statistics)}, 58(2):211--224.

\bibitem[Yin and Yuan, 2009b]{yin2009latent}
Yin, G. and Yuan, Y. (2009b).
\newblock A latent contingency table approach to dose finding for combinations
  of two agents.
\newblock {\em Biometrics}, 65(3):866--875.

\bibitem[Yuan et~al., 2016]{yuan2016bayesian}
Yuan, Y., Hess, K.~R., Hilsenbeck, S.~G., and Gilbert, M.~R. (2016).
\newblock Bayesian optimal interval design: a simple and well-performing design
  for phase i oncology trials.
\newblock {\em Clinical Cancer Research}, 22(17):4291--4301.

\end{thebibliography}

\clearpage

\begin{landscape}
\begin{table}[]
\centering
\caption{Explored Design Parameters}
\label{alter}
\resizebox{\textwidth}{!}{%
\begin{tabular}{@{}cccccc@{}}
\toprule
\multirow{2}{*}{Design} & \multirow{2}{*}{Parameter} & \multicolumn{2}{c}{Main Setting} & \multicolumn{2}{c}{Alternative Setting} \\ \cmidrule(l){3-6} 
 &  & $5 \times 3$ & $4 \times 4$ & $5 \times 3$ & $4 \times 4$ \\ \midrule
I2D & \multirow{3}{*}{$p_j$ and $q_k$} & \multirow{3}{*}{\begin{tabular}[c]{@{}c@{}}$p_j$: 0.1, 0.2, 0.25, 0.3, 0.35\\ $q_k$: 0.1, 0.3, 0.35\end{tabular}} & \multirow{3}{*}{\begin{tabular}[c]{@{}c@{}}$p_j$: 0.1, 0.2, 0.25, 0.3\\ $q_k$: 0.1, 0.2, 0.25, 0.3\end{tabular}} & \multirow{3}{*}{\begin{tabular}[c]{@{}c@{}}$p_j$: 0.05, 0.1, 0.2, 0.25, 0.3\\ $q_k$: 0.1, 0.2, 0.25\end{tabular}} & \multirow{3}{*}{\begin{tabular}[c]{@{}c@{}}$p_j$: 0.05, 0.1, 0.2, 0.22\\ $q_k$: 0.05, 0.1, 0.2, 0.22\end{tabular}} \\
Copula &  &  &  &  &  \\
DFCOMB &  &  &  &  &  \\ \midrule
POCRM & \multirow{2}{*}{Skeleton setting} & \multirow{2}{*}{\begin{tabular}[c]{@{}c@{}}half width: 0.05\\ MTD position: 11\end{tabular}} & \multirow{2}{*}{\begin{tabular}[c]{@{}c@{}}half width: 0.05\\ MTD position: 12\end{tabular}} & \multirow{2}{*}{\begin{tabular}[c]{@{}c@{}}half width: 0.03\\ MTD position: 13\end{tabular}} & \multirow{2}{*}{\begin{tabular}[c]{@{}c@{}}half width: 0.03\\ MTD position: 15\end{tabular}} \\
bCRM &  &  &  &  &  \\ \midrule
DFCOMB & \multirow{2}{*}{\begin{tabular}[c]{@{}c@{}}Escalation/de-escalation \\ probability cutoff\end{tabular}} & \multicolumn{2}{c}{\multirow{2}{*}{0.85 and 0.45}} & \multicolumn{2}{c}{\multirow{2}{*}{0.6 and 0.6}} \\
bCRM &  & \multicolumn{2}{c}{} & \multicolumn{2}{c}{} \\ \midrule
Hierarchy & \multirow{2}{*}{$\pi_{1k}$ and $\pi_{j1}$} & \multicolumn{2}{c}{\multirow{2}{*}{truth of each scenario}} & \multicolumn{2}{c}{\multirow{2}{*}{incorrect guess}} \\
gCRM &  & \multicolumn{2}{c}{} & \multicolumn{2}{c}{} \\ \bottomrule
\end{tabular}%
}
\end{table}
\end{landscape}

\clearpage

\begin{table}
\centering
\caption{Simulation Settings}
\label{setting}
\resizebox{\textwidth}{!}{%
\begin{tabular}{@{}clllllllllllllllll@{}}
\toprule
\multirow{2}{*}{Agent B} & \multicolumn{17}{c}{Agent A} \\ \cmidrule(l){2-18} 
 & \multicolumn{1}{c}{1} & \multicolumn{1}{c}{2} & \multicolumn{1}{c}{3} & \multicolumn{1}{c}{4} & \multicolumn{1}{c}{5} & \multicolumn{1}{c}{} & \multicolumn{1}{c}{1} & \multicolumn{1}{c}{2} & \multicolumn{1}{c}{3} & \multicolumn{1}{c}{4} & \multicolumn{1}{c}{5} & \multicolumn{1}{c}{} & \multicolumn{1}{c}{1} & \multicolumn{1}{c}{2} & \multicolumn{1}{c}{3} & \multicolumn{1}{c}{4} & \multicolumn{1}{c}{5} \\ \cmidrule(l){2-18} 
 & \multicolumn{5}{c}{Scenario 1} &  & \multicolumn{5}{c}{Scenario 2} &  & \multicolumn{5}{c}{Scenario 3} \\
1 & 0.05 & 0.1 & 0.15 & \textbf{0.3} & 0.45 &  & 0.15 & \textbf{0.3} & 0.45 & 0.5 & 0.6 &  & 0.02 & 0.07 & 0.1 & 0.15 & \textbf{0.3} \\
2 & 0.1 & 0.15 & \textbf{0.3} & 0.45 & 0.55 &  & \textbf{0.3} & 0.45 & 0.5 & 0.6 & 0.75 &  & 0.07 & 0.1 & 0.15 & \textbf{0.3} & 0.45 \\
3 & 0.15 & \textbf{0.3} & 0.45 & 0.5 & 0.6 &  & 0.45 & 0.55 & 0.6 & 0.7 & 0.8 &  & 0.1 & 0.15 & \textbf{0.3} & 0.45 & 0.55 \\
\multicolumn{1}{l}{} &  &  &  &  &  &  &  &  &  &  &  &  &  &  &  &  &  \\
 & \multicolumn{5}{c}{Scenario 4} &  & \multicolumn{5}{c}{Scenario 5} & \multicolumn{1}{c}{} & \multicolumn{5}{c}{Scenario 6} \\
1 & \textbf{0.3} & 0.45 & 0.6 & 0.7 & 0.8 &  & 0.01 & 0.02 & 0.08 & 0.1 & 0.11 &  & 0.05 & 0.08 & 0.1 & 0.13 & 0.15 \\
2 & 0.45 & 0.55 & 0.65 & 0.75 & 0.85 &  & 0.03 & 0.05 & 0.1 & 0.13 & 0.15 &  & 0.09 & 0.12 & 0.15 & \textbf{0.3} & 0.45 \\
3 & 0.5 & 0.6 & 0.7 & 0.8 & 0.9 &  & 0.07 & 0.09 & 0.12 & 0.15 & \textbf{0.3} &  & 0.15 & \textbf{0.3} & 0.45 & 0.5 & 0.6 \\
\multicolumn{1}{l}{} &  &  &  &  &  &  &  &  &  &  &  &  &  &  &  &  &  \\
 & \multicolumn{5}{c}{Scenario 7} & \multicolumn{1}{c}{} & \multicolumn{5}{c}{Scenario 8} & \multicolumn{1}{c}{} & \multicolumn{5}{c}{Scenario 9} \\
1 & 0.07 & 0.1 & 0.12 & 0.15 & \textbf{0.3} &  & 0.02 & 0.1 & 0.15 & 0.5 & 0.6 &  & 0.005 & 0.01 & 0.02 & 0.04 & 0.07 \\
2 & 0.15 & \textbf{0.3} & 0.45 & 0.52 & 0.6 &  & 0.05 & 0.12 & \textbf{0.3} & 0.55 & 0.7 &  & 0.02 & 0.05 & 0.08 & 0.12 & 0.15 \\
3 & \textbf{0.3} & 0.5 & 0.6 & 0.65 & 0.75 &  & 0.08 & 0.15 & 0.45 & 0.6 & 0.8 &  & 0.15 & \textbf{0.3} & 0.45 & 0.55 & 0.65 \\
\multicolumn{1}{l}{} &  &  &  &  &  &  &  &  &  &  &  &  &  &  &  &  &  \\
 & \multicolumn{5}{c}{Scenario 10} & \multicolumn{1}{c}{} & \multicolumn{4}{c}{Scenario 11} & \multicolumn{1}{c}{} & \multicolumn{1}{c}{} & \multicolumn{4}{c}{Scenario 12} &  \\
1 & 0.05 & 0.1 & 0.15 & \textbf{0.3} & 0.45 &  & 0.08 & 0.14 & 0.19 & \textbf{0.3} &  &  & 0.05 & 0.1 & 0.2 & \textbf{0.3} &  \\
2 & 0.45 & 0.5 & 0.6 & 0.65 & 0.7 &  & 0.1 & 0.2 & \textbf{0.3} & 0.55 &  &  & 0.08 & \textbf{0.3} & 0.45 & 0.5 &  \\
3 & 0.7 & 0.75 & 0.8 & 0.85 & 0.9 &  & 0.15 & \textbf{0.3} & 0.52 & 0.6 &  &  & 0.15 & 0.35 & 0.5 & 0.55 &  \\
4 &  &  &  &  &  &  & \textbf{0.3} & 0.5 & 0.6 & 0.7 &  &  & \textbf{0.3} & 0.5 & 0.6 & 0.7 &  \\
\multicolumn{1}{l}{} &  &  &  &  &  &  &  &  &  &  &  &  &  &  &  &  &  \\
 & \multicolumn{4}{c}{Scenario 13} & \multicolumn{1}{c}{} & \multicolumn{1}{c}{} & \multicolumn{4}{c}{Scenario 14} & \multicolumn{1}{c}{} & \multicolumn{1}{c}{} & \multicolumn{4}{c}{Scenario 15} &  \\
1 & 0.05 & 0.08 & 0.1 & \textbf{0.3} &  &  & 0.01 & 0.05 & 0.1 & \textbf{0.3} &  &  & 0.01 & 0.1 & 0.15 & 0.45 &  \\
2 & 0.08 & 0.1 & 0.2 & 0.35 &  &  & 0.05 & 0.1 & 0.45 & 0.5 &  &  & 0.03 & \textbf{0.3} & 0.4 & 0.5 &  \\
3 & 0.1 & 0.2 & \textbf{0.3} & 0.4 &  &  & 0.1 & 0.45 & 0.5 & 0.6 &  &  & 0.05 & 0.5 & 0.55 & 0.65 &  \\
4 & \textbf{0.3} & 0.35 & 0.4 & 0.6 &  &  & \textbf{0.3} & 0.5 & 0.6 & 0.65 &  &  & 0.08 & 0.55 & 0.6 & 0.75 &  \\ \bottomrule
\end{tabular}%
}
\end{table}

\clearpage

\begin{table}[]
\centering
\caption{Performance of MTD Selection of Designs across Scenarios when Maximum Sample Size Is 60}
\label{mtd}
\resizebox{1.1\textwidth}{!}{%
\begin{threeparttable}
\begin{tabular}{@{}lccccccccccccccc@{}}
\toprule
 & \multicolumn{15}{c}{Simulation Scenario} \\ \midrule
\multicolumn{1}{c}{Design} & \multicolumn{1}{c}{1} & \multicolumn{1}{c}{2} & \multicolumn{1}{c}{3} & \multicolumn{1}{c}{4} & \multicolumn{1}{c}{5} & \multicolumn{1}{c}{6} & \multicolumn{1}{c}{7} & \multicolumn{1}{c}{8} & \multicolumn{1}{c}{9} & \multicolumn{1}{c}{10} & \multicolumn{1}{c}{11} & \multicolumn{1}{c}{12} & \multicolumn{1}{c}{13} & \multicolumn{1}{c}{14} & \multicolumn{1}{c}{15} \\ \cmidrule(l){2-16} 
 & \multicolumn{15}{c}{Selection of correct MTD $(S_{C})$} \\ \cmidrule(l){2-16} 
I2D & {\color[HTML]{FE0000} 0.43} & 0.71 & {\color[HTML]{FE0000} 0.41} & 0.9 & 0.86 & 0.34 & 0.59 & {\color[HTML]{FE0000} 0.11} & {\color[HTML]{FE0000} 0.15} & 0.43 & 0.47 & 0.47 & {\color[HTML]{333333} 0.23} & 0.25 & {\color[HTML]{FE0000} 0.24} \\
I2D.pq & 0.58 & 0.73 & 0.54 & 0.81 & 0.88 & 0.35 & 0.54 & {\color[HTML]{FE0000} 0.2} & {\color[HTML]{FE0000} 0.09} & 0.3 & 0.58 & 0.45 & {\color[HTML]{333333} 0.28} & 0.4 & {\color[HTML]{FE0000} 0.16} \\
Copula & 0.53 & 0.6 & 0.52 & {\color[HTML]{FE0000} 0.1} & 0.9 & 0.44 & 0.65 & {\color[HTML]{FE0000} 0.15} & 0.32 & 0.26 & 0.56 & 0.45 & {\color[HTML]{FE0000} 0.12} & {\color[HTML]{FE0000} 0.12} & 0.42 \\
Copula.pq & 0.59 & 0.6 & 0.58 & {\color[HTML]{FE0000} 0.06} & 0.91 & 0.37 & 0.55 & 0.4 & {\color[HTML]{FE0000} 0.12} & {\color[HTML]{FE0000} 0.01} & 0.53 & {\color[HTML]{FE0000} 0.26} & 0.4 & {\color[HTML]{FE0000} 0.05} & {\color[HTML]{FE0000} 0.13} \\
Hierarchy & 0.61 & 0.64 & 0.62 & 0.61 & 0.81 & 0.45 & {\color[HTML]{FE0000} 0.45} & 0.3 & 0.47 & 0.52 & 0.6 & {\color[HTML]{FE0000} 0.28} & 0.47 & 0.24 & 0.52 \\
Hierarchy.pi & 0.61 & {\color[HTML]{FE0000} 0.5} & 0.71 & {\color[HTML]{FE0000} 0.32} & 0.78 & 0.42 & 0.52 & 0.42 & {\color[HTML]{FE0000} 0.09} & {\color[HTML]{FE0000} 0.11} & 0.6 & {\color[HTML]{FE0000} 0.24} & 0.39 & {\color[HTML]{FE0000} 0.11} & {\color[HTML]{FE0000} 0.17} \\
POCRM & 0.75 & 0.71 & 0.69 & 0.78 & {\color[HTML]{FE0000} 0.54} & 0.59 & 0.56 & 0.59 & 0.52 & 0.58 & 0.74 & 0.52 & 0.46 & 0.57 & 0.48 \\
POCRM.skeleton & 0.74 & 0.67 & 0.71 & 0.8 & 0.73 & 0.57 & 0.51 & 0.6 & 0.49 & 0.53 & 0.77 & 0.5 & 0.44 & 0.54 & 0.35 \\
DFCOMB & 0.54 & 0.76 & 0.66 & 0.65 & {\color[HTML]{FE0000} 0.54} & 0.33 & 0.69 & 0.48 & {\color[HTML]{FE0000} 0.15} & 0.47 & 0.44 & 0.63 & 0.37 & 0.18 & 0.67 \\
DFCOMB.pq & 0.69 & 0.8 & 0.65 & 0.64 & {\color[HTML]{FE0000} 0.52} & 0.34 & 0.71 & 0.3 & {\color[HTML]{FE0000} 0.09} & {\color[HTML]{FE0000} 0.14} & 0.57 & 0.56 & 0.33 & 0.35 & 0.45 \\
DFCOMB.cut & 0.59 & 0.79 & 0.64 & 0.65 & {\color[HTML]{FE0000} 0.54} & {\color[HTML]{FE0000} 0.26} & 0.70 & 0.36 & 0.24 & 0.57 & 0.51 & 0.63 & 0.36 & 0.15 & 0.60 \\
DFCOMB.sensitive & 0.54 & 0.78 & 0.66 & 0.62 & 0.58 & 0.33 & 0.69 & 0.44 & {\color[HTML]{FE0000} 0.15} & 0.47 & 0.46 & 0.61 & 0.38 & 0.20 & 0.64 \\
gCRM & 0.69 & 0.65 & 0.71 & {\color[HTML]{FE0000} 0.42} & 0.81 & 0.59 & 0.67 & 0.34 & 0.47 & 0.55 & 0.64 & 0.47 & 0.49 & 0.17 & 0.48 \\
gCRM.pi & 0.58 & 0.64 & 0.61 & {\color[HTML]{FE0000} 0.47} & 0.81 & 0.5 & 0.74 & 0.26 & 0.48 & 0.51 & 0.61 & 0.57 & 0.34 & 0.22 & {\color[HTML]{FE0000} 0.23} \\
cBOIN & 0.7 & 0.69 & 0.7 & 0.62 & 0.72 & 0.58 & 0.74 & 0.38 & 0.4 & 0.45 & 0.75 & 0.57 & 0.38 & 0.4 & 0.37 \\
cKeyboard & 0.67 & 0.7 & 0.7 & 0.6 & 0.72 & 0.56 & 0.71 & 0.38 & 0.4 & 0.45 & 0.73 & 0.58 & 0.38 & 0.43 & 0.36 \\
bCRM & 0.72 & 0.75 & 0.66 & 0.76 & {\color[HTML]{FE0000} 0.52} & 0.51 & 0.63 & 0.51 & 0.37 & 0.5 & 0.62 & 0.54 & 0.39 & 0.35 & 0.47 \\
bCRM.skeleton & 0.75 & 0.72 & 0.73 & 0.84 & 0.69 & 0.59 & 0.64 & 0.51 & 0.36 & 0.36 & 0.69 & 0.52 & 0.47 & 0.33 & 0.45 \\
bCRM.cut & 0.75 & 0.67 & 0.71 & 0.71 & {\color[HTML]{FE0000} 0.59} & 0.56 & 0.63 & 0.54 & 0.40 & 0.52 & 0.71 & 0.53 & 0.40 & 0.37 & 0.43 \\ \bottomrule
\end{tabular}%
\begin{tablenotes}
\item I2D: design I2D with parameter $p_j$ and $q_k$ specified in ``Main setting" of Table 1.
\item I2D.pq: design I2D with parameter $p_j$ and $q_k$ specified in ``Alternative setting".
\item Copula: design Copula with parameter $p_j$ and $q_k$ specified in ``Main setting".
\item Copula.pq: design Copula with parameter $p_j$ and $q_k$ specified in ``Alternative setting".
\item Hierarchy: design Hierarchy with $\pi_{1k}$ and $\pi_{j1}$ specified in ``Main setting".
\item Hierarchy.pi: design Hierarchy with $\pi_{1k}$ and $\pi_{j1}$ specified in ``Alternative setting".
\item POCRM: design POCRM with skeleton specified in ``Main setting".
\item POCRM.skeleton: design POCRM with skeleton specified in ``Alternative setting".
\item DFCOMB: design DFCOMB with $p_j$ and $q_k$, and escalation/de-escalation probability cutoff specified in ``Main setting".
\item DFCOMB.pq: design DFCOMB with $p_j$ and $q_k$ specified in ``Alternative setting'', and escalation/de-escalation probability cutoff specified in ``Main setting".
\item DFCOMB.cut: design DFCOMB with $p_j$ and $q_k$ specified in ``Main setting'', and escalation/de-escalation probability cutoff specified in ``Alternative setting".
\item DFCOMB.sensitive: design DFCOMB with $p_j$ and $q_k$, and escalation/de-escalation probability cutoff specified in ``Main setting", but target toxicity interval boundaries suggested by one of our reviewers.
\item gCRM: design gCRM with $\pi_{1k}$ and $\pi_{j1}$ specified in ``Main setting".
\item gCRM.pi: design gCRM with $\pi_{1k}$ and $\pi_{j1}$ specified in ``Alternative setting".
\item bCRM: design bCRM with skeleton, and escalation/de-escalation probability cutoff specified in ``Main setting".
\item bCRM.skeleton: design bCRM with skeleton specified in ``Alternative setting'', and escalation/de-escalation probability cutoff specified in ``Main setting".
\item bCRM.cut: design bCRM with skeleton specified in ``Main setting'', and escalation/de-escalation probability cutoff specified in ``Alternative setting".
\end{tablenotes}
\end{threeparttable}
}
\end{table}

\clearpage

\begin{table}[]
\centering
\caption{Performance of MTD Selection of Designs across Scenarios when Maximum Sample Size Is 60}
\label{toxicmtd}
\resizebox{1.1\textwidth}{!}{%
\begin{threeparttable}
\begin{tabular}{@{}lccccccccccccccc@{}}
\toprule
 & \multicolumn{15}{c}{Simulation Scenario} \\ \midrule
\multicolumn{1}{c}{Design} & 1 & 2 & 3 & 4 & 5 & 6 & 7 & 8 & 9 & 10 & 11 & 12 & 13 & 14 & 15 \\ \cmidrule(l){2-16} 
 & \multicolumn{15}{c}{Selection of over-toxic MTD $(S_{OT})$} \\ \cmidrule(l){2-16} 
I2D & 0.18 & 0.18 & 0.18 & 0.10 & 0 & 0.29 & 0.07 & 0.26 & 0.39 & 0.32 & 0.12 & 0.18 & 0.43 & 0.31 & 0.28 \\
I2D.pq & 0.19 & 0.19 & 0.18 & 0.19 & 0 & 0.25 & 0.17 & 0.30 & 0.42 & 0.41 & 0.19 & 0.32 & 0.43 & 0.35 & 0.45 \\
Copula & {\color[HTML]{FE0000} 0.32} & 0.19 & {\color[HTML]{FE0000} 0.36} & 0.09 & 0 & {\color[HTML]{FE0000} 0.43} & 0.22 & {\color[HTML]{FE0000} 0.58} & {\color[HTML]{FE0000} 0.55} & {\color[HTML]{FE0000} 0.63} & 0.19 & 0.43 & {\color[HTML]{FE0000} 0.66} & {\color[HTML]{FE0000} 0.69} & 0.50 \\
Copula.pq & {\color[HTML]{FE0000} 0.30} & 0.16 & {\color[HTML]{FE0000} 0.33} & 0.09 & 0 & {\color[HTML]{FE0000} 0.46} & {\color[HTML]{FE0000} 0.33} & {\color[HTML]{FE0000} 0.44} & {\color[HTML]{FE0000} 0.53} & {\color[HTML]{FE0000} 0.78} & {\color[HTML]{EDB0AC} 0.22} & {\color[HTML]{FE0000} 0.51} & 0.41 & 0.62 & {\color[HTML]{EDB0AC} 0.54} \\
Hierarchy & 0.22 & 0.18 & 0.21 & 0.16 & 0 & 0.30 & {\color[HTML]{FE0000} 0.31} & 0.30 & 0.17 & 0.20 & 0.13 & 0.37 & 0.39 & 0.41 & 0.22 \\
Hierarchy.pi & 0.22 & 0.23 & 0.14 & 0.24 & 0 & 0.29 & 0.24 & 0.27 & 0.40 & 0.39 & 0.13 & 0.43 & 0.44 & 0.54 & 0.47 \\
POCRM & 0.12 & 0.24 & 0.08 & 0.22 & 0 & 0.11 & 0.19 & 0.17 & 0.18 & 0.26 & 0.04 & 0.30 & 0.28 & 0.32 & 0.31 \\
POCRM.skeleton & 0.15 & 0.24 & 0.12 & 0.20 & 0 & 0.17 & 0.21 & 0.18 & 0.23 & 0.27 & 0.06 & 0.34 & 0.32 & 0.34 & 0.39 \\
DFCOMB & 0.18 & 0.08 & 0.17 & 0.06 & 0 & 0.27 & 0.09 & 0.27 & {\color[HTML]{FE0000} 0.57} & 0.32 & 0.08 & 0.19 & 0.25 & 0.61 & 0.21 \\
DFCOMB.pq & 0.10 & 0.06 & 0.14 & 0.08 & 0 & 0.31 & 0.13 & 0.37 & {\color[HTML]{FE0000} 0.62} & {\color[HTML]{FE0000} 0.52} & 0.08 & 0.30 & 0.28 & 0.51 & 0.45 \\
DFCOMB.cut & 0.08 & 0.07 & 0.11 & 0.08 & 0 & 0.07 & 0.05 & 0.19 & 0.37 & 0.20 & 0.02 & 0.12 & 0.16 & 0.53 & 0.23 \\
DFCOMB.sensitive & 0.25 & 0.10 & 0.21 & 0.09 & 0 & 0.36 & 0.13 & 0.36 & {\color[HTML]{FE0000} 0.65} & 0.37 & 0.12 & 0.25 & 0.32 & 0.66 & 0.27 \\
gCRM & 0.15 & 0.13 & 0.13 & 0.10 & 0 & 0.18 & 0.10 & 0.31 & 0.11 & 0.33 & 0.06 & 0.32 & 0.31 & 0.52 & 0.29 \\
gCRM.pi & 0.18 & 0.14 & 0.17 & 0.10 & 0 & 0.19 & 0.08 & 0.29 & 0.13 & 0.26 & 0.07 & 0.22 & 0.38 & 0.45 & 0.44 \\
cBOIN & 0.16 & 0.21 & 0.15 & 0.17 & 0 & 0.19 & 0.13 & 0.21 & 0.13 & 0.31 & 0.08 & 0.29 & 0.43 & 0.34 & 0.29 \\
cKeyboard & 0.17 & 0.21 & 0.14 & 0.17 & 0 & 0.20 & 0.14 & 0.21 & 0.12 & 0.31 & 0.09 & 0.27 & 0.43 & 0.34 & 0.30 \\
bCRM & 0.08 & 0.22 & 0.05 & 0.24 & 0 & 0.11 & 0.15 & 0.20 & 0.20 & 0.38 & 0.03 & 0.27 & 0.19 & 0.53 & 0.42 \\
bCRM.skeleton & 0.12 & 0.22 & 0.08 & 0.16 & 0 & 0.17 & 0.21 & 0.29 & 0.43 & {\color[HTML]{FE0000} 0.50} & 0.05 & 0.33 & 0.21 & 0.58 & 0.46 \\
bCRM.cut & 0.12 & {\color[HTML]{FE0000} 0.30} & 0.07 & {\color[HTML]{FE0000} 0.29} & 0 & 0.12 & 0.19 & 0.19 & 0.17 & 0.36 & 0.04 & 0.32 & 0.24 & 0.53 & 0.44 \\ \bottomrule
\end{tabular}%
\begin{tablenotes}
\item I2D: design I2D with parameter $p_j$ and $q_k$ specified in ``Main setting" of Table 1.
\item I2D.pq: design I2D with parameter $p_j$ and $q_k$ specified in ``Alternative setting".
\item Copula: design Copula with parameter $p_j$ and $q_k$ specified in ``Main setting".
\item Copula.pq: design Copula with parameter $p_j$ and $q_k$ specified in ``Alternative setting".
\item Hierarchy: design Hierarchy with $\pi_{1k}$ and $\pi_{j1}$ specified in ``Main setting".
\item Hierarchy.pi: design Hierarchy with $\pi_{1k}$ and $\pi_{j1}$ specified in ``Alternative setting".
\item POCRM: design POCRM with skeleton specified in ``Main setting".
\item POCRM.skeleton: design POCRM with skeleton specified in ``Alternative setting".
\item DFCOMB: design DFCOMB with $p_j$ and $q_k$, and escalation/de-escalation probability cutoff specified in ``Main setting".
\item DFCOMB.pq: design DFCOMB with $p_j$ and $q_k$ specified in ``Alternative setting'', and escalation/de-escalation probability cutoff specified in ``Main setting".
\item DFCOMB.cut: design DFCOMB with $p_j$ and $q_k$ specified in ``Main setting'', and escalation/de-escalation probability cutoff specified in ``Alternative setting".
\item DFCOMB.sensitive: design DFCOMB with $p_j$ and $q_k$, and escalation/de-escalation probability cutoff specified in ``Main setting", but target toxicity interval boundaries suggested by one of our reviewers.
\item gCRM: design gCRM with $\pi_{1k}$ and $\pi_{j1}$ specified in ``Main setting".
\item gCRM.pi: design gCRM with $\pi_{1k}$ and $\pi_{j1}$ specified in ``Alternative setting".
\item bCRM: design bCRM with skeleton, and escalation/de-escalation probability cutoff specified in ``Main setting".
\item bCRM.skeleton: design bCRM with skeleton specified in ``Alternative setting'', and escalation/de-escalation probability cutoff specified in ``Main setting".
\item bCRM.cut: design bCRM with skeleton specified in ``Main setting'', and escalation/de-escalation probability cutoff specified in ``Alternative setting".
\end{tablenotes}
\end{threeparttable}
}
\end{table}

\clearpage

\begin{table}[]
\centering
\caption{Performance of Patient Assignment of Designs across Scenarios when Maximum Sample Size Is 60}
\label{patMTD}
\resizebox{1.1\textwidth}{!}{%
\begin{threeparttable}
\begin{tabular}{@{}llllllllllllllll@{}}
\toprule
 & \multicolumn{15}{c}{Simulation Scenario} \\ \midrule
\multicolumn{1}{c}{Design} & \multicolumn{1}{c}{1} & \multicolumn{1}{c}{2} & \multicolumn{1}{c}{3} & \multicolumn{1}{c}{4} & \multicolumn{1}{c}{5} & \multicolumn{1}{c}{6} & \multicolumn{1}{c}{7} & \multicolumn{1}{c}{8} & \multicolumn{1}{c}{9} & \multicolumn{1}{c}{10} & \multicolumn{1}{c}{11} & \multicolumn{1}{c}{12} & \multicolumn{1}{c}{13} & \multicolumn{1}{c}{14} & \multicolumn{1}{c}{15} \\ \cmidrule(l){2-16} 
 & \multicolumn{15}{c}{Patient receiving correct MTD during trials $(A_{C})$} \\ \cmidrule(l){2-16} 
I2D & 0.38 & 0.57 & 0.45 & 0.71 & 0.58 & 0.23 & 0.50 & {\color[HTML]{FE0000} 0.07} & {\color[HTML]{FE0000} 0.05} & 0.33 & 0.44 & 0.39 & 0.25 & 0.27 & 0.18 \\
I2D.pq & 0.45 & 0.56 & 0.44 & {\color[HTML]{FE0000} 0.55} & 0.61 & 0.26 & 0.42 & 0.15 & {\color[HTML]{FE0000} 0.03} & 0.28 & 0.50 & 0.34 & 0.27 & 0.28 & {\color[HTML]{EDB0AC} 0.11} \\
Copula & 0.33 & 0.57 & {\color[HTML]{EDB0AC} 0.31} & 0.72 & 0.46 & 0.29 & 0.41 & 0.12 & 0.31 & 0.09 & 0.32 & 0.33 & {\color[HTML]{FE0000} 0.08} & {\color[HTML]{EDB0AC} 0.07} & 0.34 \\
Copula.pq & 0.41 & 0.55 & 0.37 & 0.63 & 0.49 & 0.27 & 0.43 & 0.26 & 0.18 & {\color[HTML]{FE0000} 0.00} & 0.40 & {\color[HTML]{EDB0AC} 0.23} & 0.27 & {\color[HTML]{EDB0AC} 0.07} & 0.12 \\
Hierarchy & 0.42 & 0.50 & 0.43 & 0.68 & 0.68 & 0.31 & 0.35 & 0.22 & 0.30 & 0.35 & 0.43 & 0.27 & 0.38 & 0.15 & 0.39 \\
Hierarchy.pi & 0.42 & {\color[HTML]{EDB0AC} 0.40} & 0.52 & {\color[HTML]{FE0000} 0.49} & 0.65 & 0.30 & 0.41 & 0.26 & {\color[HTML]{FE0000} 0.07} & {\color[HTML]{FE0000} 0.04} & 0.43 & {\color[HTML]{EDB0AC} 0.23} & 0.27 & 0.10 & 0.16 \\
POCRM & 0.53 & 0.53 & 0.47 & 0.64 & 0.33 & 0.37 & 0.43 & 0.35 & 0.30 & 0.36 & 0.57 & 0.39 & 0.33 & 0.36 & 0.33 \\
POCRM.skeleton & 0.52 & 0.48 & 0.48 & 0.68 & 0.47 & 0.36 & 0.38 & 0.35 & 0.29 & 0.33 & 0.56 & 0.38 & 0.34 & 0.35 & 0.20 \\
DFCOMB & {\color[HTML]{FE0000} 0.23} & 0.45 & 0.37 & 0.91 & 0.39 & {\color[HTML]{FE0000} 0.16} & {\color[HTML]{EDB0AC} 0.33} & 0.23 & 0.13 & 0.16 & {\color[HTML]{FE0000} 0.20} & 0.28 & {\color[HTML]{FE0000} 0.12} & 0.11 & 0.35 \\
DFCOMB.pq & 0.34 & 0.48 & 0.37 & 0.91 & 0.37 & {\color[HTML]{FE0000} 0.17} & {\color[HTML]{EDB0AC} 0.33} & 0.20 & {\color[HTML]{FE0000} 0.06} & {\color[HTML]{FE0000} 0.04} & 0.28 & 0.29 & 0.19 & 0.23 & 0.28 \\
DFCOMB.cut & 0.34 & 0.46 & 0.39 & 0.77 & 0.51 & 0.23 & 0.41 & 0.17 & 0.17 & 0.28 & 0.39 & 0.33 & 0.21 & 0.11 & 0.26 \\
DFCOMB.sensitive & {\color[HTML]{FE0000} 0.23} & 0.45 & 0.37 & 0.91 & 0.39 & {\color[HTML]{FE0000} 0.16} & {\color[HTML]{EDB0AC} 0.33} & 0.23 & 0.13 & 0.16 & {\color[HTML]{FE0000} 0.20} & 0.28 & {\color[HTML]{FE0000} 0.12} & 0.11 & 0.35 \\
gCRM.b & 0.46 & 0.45 & 0.49 & 0.75 & 0.69 & 0.36 & 0.46 & 0.27 & 0.29 & 0.36 & 0.44 & 0.34 & 0.34 & 0.14 & 0.33 \\
gCRM.pi & 0.37 & 0.44 & 0.41 & 0.75 & 0.68 & 0.32 & 0.54 & 0.16 & 0.25 & 0.31 & 0.43 & 0.40 & 0.25 & 0.18 & 0.18 \\
cBOIN & 0.43 & 0.49 & 0.40 & 0.72 & 0.43 & 0.34 & 0.46 & 0.21 & 0.26 & 0.20 & 0.44 & 0.37 & 0.23 & 0.21 & 0.25 \\
cKeyboard & 0.42 & 0.49 & 0.40 & 0.72 & 0.43 & 0.33 & 0.44 & 0.21 & 0.25 & 0.20 & 0.43 & 0.37 & 0.23 & 0.22 & 0.24 \\
bCRM & 0.43 & 0.52 & 0.37 & 0.70 & {\color[HTML]{FE0000} 0.24} & 0.26 & 0.35 & 0.24 & 0.18 & 0.24 & 0.36 & 0.37 & 0.26 & 0.21 & 0.30 \\
bCRM.skeleton & 0.46 & 0.49 & 0.44 & 0.77 & 0.34 & 0.28 & 0.37 & 0.25 & 0.16 & 0.19 & 0.42 & 0.37 & 0.31 & 0.20 & 0.28 \\
bCRM.cut & 0.47 & 0.44 & 0.47 & {\color[HTML]{FE0000} 0.52} & 0.36 & 0.30 & 0.35 & 0.26 & 0.15 & 0.25 & 0.47 & 0.35 & 0.29 & 0.21 & 0.24 \\ \bottomrule
\end{tabular}%
\begin{tablenotes}
\item I2D: design I2D with parameter $p_j$ and $q_k$ specified in ``Main setting" of Table 1.
\item I2D.pq: design I2D with parameter $p_j$ and $q_k$ specified in ``Alternative setting".
\item Copula: design Copula with parameter $p_j$ and $q_k$ specified in ``Main setting".
\item Copula.pq: design Copula with parameter $p_j$ and $q_k$ specified in ``Alternative setting".
\item Hierarchy: design Hierarchy with $\pi_{1k}$ and $\pi_{j1}$ specified in ``Main setting".
\item Hierarchy.pi: design Hierarchy with $\pi_{1k}$ and $\pi_{j1}$ specified in ``Alternative setting".
\item POCRM: design POCRM with skeleton specified in ``Main setting".
\item POCRM.skeleton: design POCRM with skeleton specified in ``Alternative setting".
\item DFCOMB: design DFCOMB with $p_j$ and $q_k$, and escalation/de-escalation probability cutoff specified in ``Main setting".
\item DFCOMB.pq: design DFCOMB with $p_j$ and $q_k$ specified in ``Alternative setting'', and escalation/de-escalation probability cutoff specified in ``Main setting".
\item DFCOMB.cut: design DFCOMB with $p_j$ and $q_k$ specified in ``Main setting'', and escalation/de-escalation probability cutoff specified in ``Alternative setting".
\item DFCOMB.sensitive: design DFCOMB with $p_j$ and $q_k$, and escalation/de-escalation probability cutoff specified in ``Main setting", but target toxicity interval boundaries suggested by one of our reviewers.
\item gCRM: design gCRM with $\pi_{1k}$ and $\pi_{j1}$ specified in ``Main setting".
\item gCRM.pi: design gCRM with $\pi_{1k}$ and $\pi_{j1}$ specified in ``Alternative setting".
\item bCRM: design bCRM with skeleton, and escalation/de-escalation probability cutoff specified in ``Main setting".
\item bCRM.skeleton: design bCRM with skeleton specified in ``Alternative setting'', and escalation/de-escalation probability cutoff specified in ``Main setting".
\item bCRM.cut: design bCRM with skeleton specified in ``Main setting'', and escalation/de-escalation probability cutoff specified in ``Alternative setting".
\end{tablenotes}
\end{threeparttable}
}
\end{table}

\clearpage

\begin{table}[]
\centering
\caption{Performance of Patient Assignment of Designs across Scenarios when Maximum Sample Size Is 60}
\label{patToxic}
\resizebox{1.1\textwidth}{!}{%
\begin{threeparttable}
\begin{tabular}{@{}llllllllllllllll@{}}
\toprule
 & \multicolumn{15}{c}{Simulation Scenario} \\ \midrule
\multicolumn{1}{c}{Design} & \multicolumn{1}{c}{1} & \multicolumn{1}{c}{2} & \multicolumn{1}{c}{3} & \multicolumn{1}{c}{4} & \multicolumn{1}{c}{5} & \multicolumn{1}{c}{6} & \multicolumn{1}{c}{7} & \multicolumn{1}{c}{8} & \multicolumn{1}{c}{9} & \multicolumn{1}{c}{10} & \multicolumn{1}{c}{11} & \multicolumn{1}{c}{12} & \multicolumn{1}{c}{13} & \multicolumn{1}{c}{14} & \multicolumn{1}{c}{15} \\ \cmidrule(l){2-16} 
 & \multicolumn{15}{c}{Patient receiving over-toxic doses during trials $(A_{OT})$} \\ \cmidrule(l){2-16} 
I2D & 0.29 & 0.27 & 0.19 & 0.29 & 0 & 0.26 & 0.16 & 0.37 & 0.35 & 0.31 & 0.15 & 0.28 & 0.36 & 0.38 & 0.42 \\
I2D.pq & 0.26 & 0.35 & 0.22 & {\color[HTML]{FE0000} 0.45} & 0 & 0.30 & 0.25 & 0.33 & 0.36 & 0.34 & 0.20 & 0.38 & 0.40 & 0.44 & {\color[HTML]{EDB0AC} 0.55} \\
Copula & 0.17 & 0.13 & 0.19 & 0.28 & 0 & 0.18 & 0.13 & 0.31 & 0.23 & 0.41 & 0.13 & 0.28 & 0.39 & 0.44 & 0.30 \\
Copula.pq & 0.17 & 0.20 & 0.16 & 0.37 & 0 & 0.18 & 0.17 & 0.26 & 0.23 & 0.45 & 0.14 & 0.36 & 0.27 & 0.44 & 0.41 \\
Hierarchy & {\color[HTML]{EDB0AC} 0.35} & 0.35 & {\color[HTML]{EDB0AC} 0.31} & 0.32 & 0 & {\color[HTML]{EDB0AC} 0.36} & {\color[HTML]{EDB0AC} 0.36} & {\color[HTML]{EDB0AC} 0.43} & 0.30 & 0.44 & {\color[HTML]{FE0000} 0.24} & 0.44 & 0.44 & {\color[HTML]{EDB0AC} 0.60} & 0.38 \\
Hierarchy.pi & {\color[HTML]{EDB0AC} 0.34} & {\color[HTML]{FE0000} 0.46} & 0.23 & {\color[HTML]{FE0000} 0.51} & 0 & 0.33 & 0.33 & 0.38 & 0.38 & {\color[HTML]{EDB0AC} 0.52} & {\color[HTML]{FE0000} 0.24} & {\color[HTML]{EDB0AC} 0.48} & {\color[HTML]{EDB0AC} 0.47} & 0.58 & 0.51 \\
POCRM & 0.16 & 0.32 & 0.11 & 0.36 & 0 & 0.15 & 0.24 & 0.24 & 0.21 & 0.38 & 0.10 & 0.33 & 0.25 & 0.36 & 0.34 \\
POCRM.skeleton & 0.20 & 0.37 & 0.16 & 0.29 & 0 & 0.21 & 0.27 & 0.29 & 0.26 & 0.40 & 0.12 & 0.34 & 0.29 & 0.38 & 0.45 \\
DFCOMB & 0.11 & 0.05 & 0.14 & 0.09 & 0 & 0.14 & 0.08 & 0.23 & 0.25 & 0.23 & 0.06 & 0.15 & 0.21 & 0.45 & 0.26 \\
DFCOMB.pq & 0.09 & 0.07 & 0.14 & 0.09 & 0 & 0.17 & 0.13 & 0.29 & 0.28 & 0.33 & 0.06 & 0.18 & 0.19 & 0.36 & 0.30 \\
DFCOMB.cut & 0.21 & 0.17 & 0.24 & 0.23 & 0 & 0.22 & 0.19 & 0.35 & {\color[HTML]{EDB0AC} 0.42} & 0.30 & 0.10 & 0.24 & 0.28 & 0.55 & 0.44 \\
DFCOMB.sensitive & 0.11 & 0.05 & 0.14 & 0.09 & 0 & 0.14 & 0.08 & 0.23 & 0.25 & 0.23 & 0.06 & 0.15 & 0.21 & 0.45 & 0.26 \\
gCRM.b & 0.27 & 0.27 & 0.24 & 0.25 & 0 & 0.29 & 0.21 & 0.34 & 0.23 & 0.42 & 0.16 & 0.37 & 0.38 & 0.54 & 0.35 \\
gCRM.pi & 0.31 & 0.28 & 0.27 & 0.25 & 0 & 0.28 & 0.17 & 0.36 & 0.23 & 0.37 & 0.17 & 0.31 & 0.41 & 0.47 & 0.46 \\
cBOIN & 0.20 & 0.27 & 0.17 & 0.28 & 0 & 0.22 & 0.20 & 0.27 & 0.21 & 0.38 & 0.15 & 0.28 & 0.33 & 0.37 & 0.32 \\
cKeyboard & 0.20 & 0.27 & 0.17 & 0.28 & 0 & 0.22 & 0.21 & 0.27 & 0.20 & 0.38 & 0.15 & 0.28 & 0.33 & 0.37 & 0.32 \\
bCRM & 0.11 & 0.27 & 0.06 & 0.30 & 0 & 0.11 & 0.17 & 0.23 & 0.16 & 0.33 & 0.07 & 0.21 & 0.15 & 0.40 & 0.33 \\
bCRM.skeleton & 0.15 & 0.26 & 0.08 & 0.23 & 0 & 0.14 & 0.19 & 0.27 & 0.22 & 0.34 & 0.08 & 0.24 & 0.18 & 0.43 & 0.36 \\
bCRM.cut & 0.23 & {\color[HTML]{FE0000} 0.45} & 0.12 & {\color[HTML]{FE0000} 0.48} & 0 & 0.19 & 0.30 & 0.34 & 0.28 & 0.45 & 0.13 & 0.35 & 0.23 & 0.50 & 0.47 \\ \bottomrule
\end{tabular}%
\begin{tablenotes}
\item I2D: design I2D with parameter $p_j$ and $q_k$ specified in ``Main setting" of Table 1.
\item I2D.pq: design I2D with parameter $p_j$ and $q_k$ specified in ``Alternative setting".
\item Copula: design Copula with parameter $p_j$ and $q_k$ specified in ``Main setting".
\item Copula.pq: design Copula with parameter $p_j$ and $q_k$ specified in ``Alternative setting".
\item Hierarchy: design Hierarchy with $\pi_{1k}$ and $\pi_{j1}$ specified in ``Main setting".
\item Hierarchy.pi: design Hierarchy with $\pi_{1k}$ and $\pi_{j1}$ specified in ``Alternative setting".
\item POCRM: design POCRM with skeleton specified in ``Main setting".
\item POCRM.skeleton: design POCRM with skeleton specified in ``Alternative setting".
\item DFCOMB: design DFCOMB with $p_j$ and $q_k$, and escalation/de-escalation probability cutoff specified in ``Main setting".
\item DFCOMB.pq: design DFCOMB with $p_j$ and $q_k$ specified in ``Alternative setting'', and escalation/de-escalation probability cutoff specified in ``Main setting".
\item DFCOMB.cut: design DFCOMB with $p_j$ and $q_k$ specified in ``Main setting'', and escalation/de-escalation probability cutoff specified in ``Alternative setting".
\item DFCOMB.sensitive: design DFCOMB with $p_j$ and $q_k$, and escalation/de-escalation probability cutoff specified in ``Main setting", but target toxicity interval boundaries suggested by one of our reviewers.
\item gCRM: design gCRM with $\pi_{1k}$ and $\pi_{j1}$ specified in ``Main setting".
\item gCRM.pi: design gCRM with $\pi_{1k}$ and $\pi_{j1}$ specified in ``Alternative setting".
\item bCRM: design bCRM with skeleton, and escalation/de-escalation probability cutoff specified in ``Main setting".
\item bCRM.skeleton: design bCRM with skeleton specified in ``Alternative setting'', and escalation/de-escalation probability cutoff specified in ``Main setting".
\item bCRM.cut: design bCRM with skeleton specified in ``Main setting'', and escalation/de-escalation probability cutoff specified in ``Alternative setting".
\end{tablenotes}
\end{threeparttable}
}
\end{table}

\clearpage

\end{document}